\RequirePackage{fix-cm}
\documentclass[twocolumn,aps,superscriptaddress,nofootinbib,floatfix,linenumbers]{revtex4}
\usepackage{bm}
\usepackage{graphicx} 
\usepackage[utf8]{inputenc}
\usepackage[dvipsnames]{xcolor}
\usepackage{amsmath}
\usepackage{amssymb}
\usepackage{tabularx}
\usepackage{booktabs}
\usepackage{siunitx}
\usepackage{multirow}
\usepackage{amsfonts}
\usepackage[pagewise]{lineno}
\usepackage{graphics}
\usepackage{float}
\usepackage{placeins}
\usepackage[hidelinks]{hyperref}
\usepackage{color,amsmath,amssymb,bm}
\usepackage{tensor}
\usepackage[normalem]{ulem} 
\usepackage{hyphenat}
\usepackage{microtype}

\AtBeginDocument{%
  \heavyrulewidth=.08em
  \lightrulewidth=.05em
  \cmidrulewidth=.03em
  \belowrulesep=.65ex
  \belowbottomsep=0pt
  \aboverulesep=.4ex
  \abovetopsep=0pt
  \cmidrulesep=\doublerulesep
  \cmidrulekern=.5em
  \defaultaddspace=.5em
}

\newcommand{\invRe}{\text{Re}^{-1}}

\begin{document}

\title{Far-from-equilibrium attractors with Full Relativistic Boltzmann approach in 3+1D: moments of distribution function and anisotropic flows $v_n$}

\author{Vincenzo Nugara}
\email{vincenzo.nugara@phd.unict.it}
\affiliation{Department of Physics and Astronomy, University of Catania, Via S. Sofia 64, I-95125 Catania}
\affiliation{INFN-Laboratori Nazionali del Sud, Via S. Sofia 62, I-95123 Catania, Italy}

\author{Vincenzo Greco}
\affiliation{Department of Physics and Astronomy, University of Catania, Via S. Sofia 64, I-95125 Catania}
\affiliation{INFN-Laboratori Nazionali del Sud, Via S. Sofia 62, I-95123 Catania, Italy}

\author{Salvatore Plumari}
\email{salvatore.plumari@dfa.unict.it}
\affiliation{Department of Physics and Astronomy, University of Catania, Via S. Sofia 64, I-95125 Catania}
\affiliation{INFN-Laboratori Nazionali del Sud, Via S. Sofia 62, I-95123 Catania, Italy}

\begin{abstract}
We employ the Full Relativistic Boltzmann Transport approach for a conformal system in 3+1D to study the universal behaviour in moments of the distribution function and anisotropic flows. We investigate different transverse system sizes $R$ and interaction strength $\eta/s$ and identify universality classes based upon the interplay between  $R$ and the mean free path; we show that each of this classes can be identified by a particular value of the opacity $\hat \gamma$, which has been previously introduced in literature.
Our results highlight that, at early times, the inverse Reynolds number and momentum moments of the distribution function display universal behaviour, converging to a 1D attractor driven by longitudinal expansion. This indicates that systems of different sizes and interaction strengths tend to approach equilibrium in a similar manner. We provide a detailed analysis of how the onset of transverse flow affects these moments at later times.
Moreover, we investigate 
the system size and $\eta/s$ dependence for the harmonic flows $v_2$, $v_3$, $v_4$ and their response functions, along with the impact of the $\eta/s$ and the system transverse size on the dissipation of initial azimuthal correlations in momentum space. 
Finally, we introduce the normalised elliptic flow $v_2/v_{2,eq}$, showing the emergence of attractor behaviour in the regime of large opacity. These results offer new insights into how different systems evolve towards equilibrium and the role that system size and interaction play in this process.
\end{abstract}

\maketitle

\section{Introduction}
Ultra-Relativistic Heavy-Ion Collisions (uRHICs) provide a valuable tool for studying the formation and evolution of the Quark-Gluon Plasma (QGP). To infer the fundamental properties of nuclear matter under extreme conditions, dynamical modelling is indispensable. The dynamics of this deconfined state of strongly interacting matter can be successfully described using both macroscopic hydrodynamics theories \cite{Teaney:2009qa, Gardim:2012yp, Romatschke:2017ejr, Florkowski:2017olj, Heinz:2013th, Denicol:2012cn} and microscopic kinetic transport approaches \cite{Ruggieri:2013bda,  Ruggieri:2015yea, Plumari:2015sia, Plumari:2019gwq, Xu:2004mz, Uphoff:2014cba, Cassing:2009vt, Bratkovskaya:2011wp, Soloveva:2021quj, Kurkela:2018vqr}. Comparisons of fluid dynamic models with experimental data on collective flows in nucleus-nucleus ($AA$) collisions have demonstrated that the created matter behaves like an almost perfect fluid, exhibiting minimal shear viscosity over entropy density ratio ($\eta/s$). Surprisingly, recent experimental data have shown significant azimuthal anisotropy persisting also in smaller collision systems, such as proton-nucleus ($pA$) and proton-proton ($pp$) collisions in high multiplicity events \cite{ALICE:2014dwt, ATLAS:2017hap, CMS:2017kcs, Loizides:2016tew}. The origin of this collective signal in experimental data remains unclear. Nevertheless, simulations of relativistic hydrodynamics, particularly in its viscous and anisotropic formulations, have proven to be powerful tools. They effectively describe the evolution of macroscopic quantities across various scenarios, even in dynamics far from equilibrium, including small collision systems like $pA$ and $pp$, raising the question of the possible formation of a small droplet of QGP in these collisions systems \cite{Yan:2013laa, Romatschke:2015gxa, Habich:2015rtj, Weller:2017tsr, Shen:2016zpp, Mantysaari:2017cni, Romatschke:2017acs, Grosse-Oetringhaus:2024bwr}. This capability highlights the robustness of hydrodynamic models in capturing the complex behaviour of nuclear matter under extreme conditions. However, it also underscores the importance of understanding the thermalisation mechanism, the evolution time scales and the role of the system size to comprehend the nature of this collective behaviour in small collision systems. In recent years, from a theoretical perspective, a new feature has emerged from the study of scaling properties of relativistic hydrodynamics and microscopic kinetic equations, revealing a universal behaviour that could account for the similarities observed in different collision systems. Initially, these studies were performed in the context of relativistic hydrodynamics and subsequently extended to microscopic transport theory, classical Yang-Mills equations, and AdS/CFT calculations \cite{Heller:2015dha, Strickland:2017kux, Chattopadhyay:2019jqj, Jaiswal:2019cju, Blaizot:2019scw, Alalawi:2020zbx, Heller:2020anv, Kurkela:2018vqr, Kurkela:2019set, Almaalol:2020rnu, Behtash:2017wqg, Blaizot:2017ucy, Strickland:2018ayk, Heller:2018qvh, Kamata:2020mka, Tanji:2017suk, Brewer:2022vkq, Ambrus:2021sjg, Berges:2013eia, Berges:2013fga, Kurkela:2019set, Heller:2011ju, Du:2023bwi}. More recently, this analysis has been extended by means of a Full Boltzmann Transport approach to study universal behaviour in the moments of the phase-space distribution function, investigating also the role of a non-boost invariant distribution and the role of a temperature dependent $\eta/s(T)$ \cite{Nugara:2023eku}.
While the initial studies primarily focused on 0+1D simulations, recent research in the realm of effective kinetic theory has extended these studies to the more realistic scenario of 3+1D simulations \cite{Ambrus:2021sjg, Ambrus:2022koq, Boguslavski:2023jvg}. On the other hand, RTA approach has been extended also to the case of a mixture of quark and gluon fluids \cite{Frasca:2024ege}.\\
Recently, studies employing an effective kinetic description in Relaxation Time Approximation (RTA) have studied more realistic scenarios considering 3+1D simulations and the role of the onset of the  transverse flow from small to large collision systems showing that at early times the dynamics of the fireball is governed by the local 0+1D Bjorken flow attractor due to the rapid expansion along the longitudinal direction \cite{Ambrus:2022koq}. Within the effective kinetic description in RTA and Isotropization Time Approximation (ITA) \cite{Kurkela:2019kip} it has been shown that the evolution of the system depends on a single dimensionless parameter, the so-called opacity $\hat \gamma$, that encodes information about the system size, the interaction of the system in terms of shear viscosity over entropy density $\eta/s$ and the initial energy density \cite{Kurkela:2018ygx, Kurkela:2018vqr, Kurkela:2019set, Ambrus:2021fej}. Studies have been performed also by means of kinetic theory approaches in case of non-equilibrated systems due to a few-scattering collision kernel, which corresponds to a small opacity limit \cite{Bachmann:2022cls}.

In this paper we extend the previous work based on Full Relativistic Boltzmann Transport approach in 1+0D and 1+1D \cite{Nugara:2023eku} to a more realistic 3+1D simulation where we study the role of the transverse dynamics for different system sizes, $\eta/s$ and initial eccentricities, and how this dynamics affect the appearance of the attractors in moments of the distribution function. Moreover, our model allows us to study the presence of attractors for the anisotropic flows $v_n = \langle \cos(n\varphi)\rangle$ as well as to analyse, {for the first time in a wide range of opacity}, the impact of an initial azimuthal momentum anisotropy quantified in terms of a finite initial $v_2$, as suggested by the Colour Glass Condensate model, with particular attention to small systems, where the initial conditions are expected to sensitively affect the final observables.\\

The paper is organised as follows.
In Sec. \ref{sec:RBT} we show in detail the Full Relativistic Boltzmann approach and we discuss the initial conditions used in our work. In Sec. \ref{sec:reynolds} we discuss the space time evolution of the Reynolds number in order to quantify the deviation of the system from equilibrium and study the emergence of universal behaviour.
In Sec. \ref{sec:moments}  we show the time evolution of the distribution function moments at fixed specific viscosity. Moreover, we discuss the role of the onset of the transverse flow. Finally, in Sec. \ref{sec:flows} we focus on the anisotropic flows $v_n$ for $n=2,3,4$ by analysing the response functions $v_n/\epsilon_n$. We also discuss the impact of initial momentum anisotropies on final anisotropic flows and introduce the normalised elliptic flow $v_2/v_{2,eq}$ in order to quantify the deviation of $v_2$ from that of a thermalised medium, delving into the search for universal scaling.
The conclusions of our work are summarised in Sec. \ref{sec:conclusions}.\newline

\section{Relativistic Boltzmann Transport Approach}
\label{sec:RBT}
In this work we employ the Relativistic Boltzmann Transport (RBT) approach used to describe the evolution of the quark-gluon plasma (QGP) matter created in uRHICs.   
In our simulation we employ a relativistic transport code developed in these years to perform studies of the QGP dynamics for different collision systems from RHIC to LHC energies 
\cite{Ferini:2008he, Plumari:2012ep, Scardina:2012mik, Puglisi:2014sha, Scardina:2014gxa, Plumari:2015sia, Plumari:2015cfa, Scardina:2017ipo, Plumari:2019gwq, Plumari:2019hzp, Sun:2019gxg, Sambataro:2020pge, Gabbana:2019uqv,Oliva:2020doe, Nugara:2023eku}.
We describe QGP matter with an on-shell one-body
distribution function $f$, depending on space-time coordinates $x^{\mu}=\left(t,\mathbf{x}\right)$
and 4-momentum $p^{\mu}=(p^0,\mathbf{p})$, being $p^{0}=E_{\mathbf{p}}\equiv\sqrt{m^{2}+\mathbf{p}^{2}}$ where the space-time evolution of $f$ is
governed by the following RBT equation
\begin{equation}
p^{\mu}\partial_{\mu}f(x,p)=C\left[f(x,p)\right]_{\mathbf{p}},
\label{eq:RBT}
\end{equation}
where $\partial_{\mu}\equiv\partial/\partial x^{\mu}$ is the gradient with respect to space-time coordinates $x^{\mu}$.
In this work the QGP bulk consists of a degenerate system of massless partons; this corresponds to simulate a one-component conformal ($m=0$) fluid.
Moreover, we consider only $2\leftrightarrow 2$ collisions, which lead to the following collision integral:
\begin{gather}
C\left[f\right]_{\mathbf{p}} =
\intop\frac{\text{d}^{3}p_{2}}{2E_{\mathbf{p}_{2}}\left(2\pi\right)^{3}}
\intop\frac{\text{d}^{3}p_{1'}}{2E_{\mathbf{p}_{1'}}\left(2\pi\right)^{3}}
\int\frac{\text{d}^{3}p_{2'}}{2E_{\mathbf{p}_{2'}}\left(2\pi\right)^{3}}\nonumber \\
\times\left(f_{1'}f_{2'}-f_{1}f_{2}\right)  \left|\mathcal{M}\right|^{2}\delta^{\left(4\right)}\left(p_{1}+p_{2}-p_{1'}-p_{2'}\right),
\end{gather}
where $f_i=f(p_i)$. $\mathcal{M}$ denotes the transition amplitude for the elastic processes which is related to the differential cross section {$|{\cal M}|^2=16 \pi\,\mathfrak{s}^2 d\sigma/d\mathfrak{t}$} with {$\mathfrak{s}$} and  $\mathfrak{t}$ being the Mandelstam variables. 
In this work, we restrict ourselves to considering collisions with isotropic cross section: the differential cross section in the matrix ${\cal M}$ is determined by the total cross section $\sigma$ through the relation $d\sigma/d\mathfrak{t}=\sigma/\mathfrak{s}$.
We determine locally in space and time the total cross section in order to fix globally $\eta/s$.
An analytical relation between $\eta$, the temperature $T(x)$, the fugacity $\Gamma(x)$\footnote{Since we deal only with elastic collisions, the number of particles is conserved and the system develops a $\Gamma(x)\ne 1$.} and total cross section $\sigma$ is obtained within the Chapman-Enskog approximation:
\begin{equation}
\sigma(x)=1.2\frac{T(x)}{n(x)\left(4-\ln{\Gamma(x)}\right)\eta/s}.
\label{eq:sigma}
\end{equation}
See Ref.\cite{Nugara:2023eku, Plumari:2019gwq,Plumari:2015cfa} for more details.
It has been shown in \cite{Gabbana:2019uqv} that this approach is able to describe the dynamical evolution in wide range of $\eta/s$ ratio, even below $\eta/s=0.05$ which is smaller than the  conjectured minimal bound of $1/4\pi$ for the QGP \cite{Kovtun:2004de}.\\

In order to extract the local thermodynamic quantities, one has to compute the energy density $\varepsilon$ and particle density $n$ in the LRF, which we choose to be the Landau LRF. We determine the energy density $\varepsilon$ and the fluid four-velocity $u_{\mu}$ by solving the
eigenvalue problem for the energy-momentum tensor, with $\varepsilon$ being the largest eigenvalue:
\begin{equation}\label{eq:Landau_frame}
    T^{\mu\nu} u_{\nu} = \varepsilon u^{\mu},
\end{equation}
where $u^\mu =\gamma\ (1,\boldsymbol{\beta})$, with $\boldsymbol{\beta}$ being the 3-velocity of the fluid element and $\gamma = 1/\sqrt{1-\boldsymbol{\beta}^2}$ the corresponding Lorentz factor. In kinetic theory $T^{\mu\nu}$ is given by
\begin{equation}
    T^{\mu\nu}(x)= \int dP \, p^{\mu} p^{\nu} f(x,p),
    \label{eq:Tmunu_kinetic}
\end{equation}
with $dP=d^3p/\left((2\pi)^3p^0\right)$ being the integration measure in the momentum space.
The local particle density is given by the expression
\begin{equation} \label{eq:densityLRF}
    n = n^\mu u_\mu,
\end{equation}
where the four-current $n^\mu$ in kinetic theory reads
\begin{equation}\label{eq:Nmu_kinetic}
    n^\mu = \int dP \, p^\mu f(x,p).
\end{equation}
We define space-time dependent temperature $T(x)$ and fugacity $\Gamma(x)$ following the Landau matching conditions as:
\begin{equation} \label{eq:matching_conditions}
    T(x) \equiv\frac{\varepsilon (x)}{3\,n(x)}, \qquad
    \Gamma(x)\equiv\frac{n(x)}{d \,T(x)^3/\pi^2},
\end{equation}
where $d$ is the number of degrees of freedom of the system, assumed to be $d=1$ henceforth. 

We solve Eq. \eqref{eq:RBT} numerically by discretising the space-time in cells and using the test particle method \cite{Wong:1982zzb}, where $f(x,p)$ is sampled by considering a finite number $N$ of point-like test particles:
\begin{equation}
f\left(t,\mathbf{x},\mathbf{p}\right)=\frac{1}{N_{\text{test}}}\sum_{i=1}^{N}\delta^{\left(3\right)}\left(\mathbf{x}-\mathbf{x}_{i}\left(t\right)\right)\delta^{\left(3\right)}\left(\mathbf{p}-\mathbf{p}_{i}\left(t\right)\right),\label{eq:f_test}
\end{equation}
where $N_{test}$ is the number of test particles per real particle and $\mathbf{x}_{i}\left(t\right)$ and $\mathbf{p}_{i}\left(t\right)$ are the position and 3-momentum of the $i-$th test particle at time $t$
respectively. 
It can be shown that Eq. \eqref{eq:f_test} is a solution of the RBT equation provided 
that test particle coordinates satisfy the relativistic Hamilton equations of motion
\begin{align}
 & \dot{\mathbf{x}}_{i}\left(t\right)=\frac{\mathbf{p}_{i}\left(t\right)}{p_{i}^0\left(t\right)}\label{eq:Particle position equation of motion}\\
 & \dot{\mathbf{p}}_{i}\left(t\right)={coll.}
 \label{eq:eq_motion}
\end{align}
These equations are solved with a 4th-order Runge-Kutta integration method on a space-time grid.
The term $coll.$ on the RHS of the Hamilton equation indicates the effect of the collision integral, which we implement through the stochastic algorithm as in Ref. \cite{Xu:2004mz,Ferini:2008he}, which requires the space-time discretisation.\\
In the transport code, space is discretised in cells with fixed size in $x$, $y$ and space-time rapidity $\eta_s$ (see Ref. \cite{Nugara:2023eku} for more details).
As far as the simulations whose results will be shown in the following, we fixed $\Delta x = 0.12$ fm, $\Delta y = 0.12 $ fm and $\Delta \eta_s = 0.25$.  Finally, we are able to simulate systems with up to $3\cdot 10^8$ total particles, achieving quite good statistics. The results shown in this paper are considering a selection at mid-rapidity with $|\eta_s|<0.125$. Using the numerical setup described above we have verified that the convergence and stability of the Relativistic Boltzmann Equation is guaranteed.\\
The code is written in C language and has been optimised for High Performance Computing (HPC). The running time is linear with the number of total particles and time steps: for values of 2000 time steps and $10^8$ total particles as used in this work, a simulation requires 100 core-hours. 

\subsection{Initial conditions}
The initial conditions of the Boltzmann equation  are specified by the
distribution function $f(x,p)$ in coordinate ($x$) and momentum
($p$) space at initial proper time $\tau_0=0.1$ fm. 
The initial distribution is uniform in space-time rapidity $\eta_s$ in the range $[-2.5,2.5]$; the initial density profile in the transverse plane is given by a gaussian $f(x_\perp)\propto \exp(-x_\perp^2/R^2)$, where $\mathbf{x}_\perp = (x,y)$ and the transverse radius $R$ determines the initial extension of the fireball.\\
Following \cite{Plumari:2015sia}, we induce eccentricities $\epsilon_n$  by deforming the initial
distribution, thus mimicking anisotropies due to the initial elliptic space shape and/or initial state fluctuations~\cite{Alver:2010gr, Plumari:2015cfa}. This is usually achieved by deforming directly the initial energy density, as in hydrodynamics or kinetic theory works ~\cite{Alver:2010dn,Retinskaya:2013gca, Kurkela:2019kip, Ambrus:2022qya}. 
In our transport simulation, instead, the initial conditions are specified
by the test particles' position and thus it is sufficient to shift by a small amount these coordinates. Introducing the complex notation
$z=x+iy$, in order to generate an anisotropy in transverse plane $\epsilon_n$ we shift $z$ according to 
\begin{equation}
\label{deformation}
z\to z+\delta z\equiv z-\alpha \bar z^{n-1},
\end{equation}
where 
$\bar z\equiv x-iy$, and $\alpha$ is a real positive quantity chosen in
such a way that the correction is small. 
As shown in \cite{Plumari:2015sia}, to leading order in $\alpha$ and
for a large number of particles, one obtains
\begin{eqnarray}\label{eq:eccentricity}
\epsilon_n &=& \dfrac{ \sqrt{ \langle x_\perp^{n} \cos(n\phi)\rangle^2 + \langle x_\perp^{n} \sin(n\phi)\rangle^2 } }{\langle x_\perp^n \rangle} \simeq \nonumber \\ 
&\simeq &n\alpha\dfrac{\langle x_\perp^{2(n-1)}\rangle}{\langle x_\perp^n\rangle}.
\label{eq:ecc_first_order}
\end{eqnarray}
{Specifically, starting with a Gaussian distribution in the transverse plane, we can analytically evaluate Eq. \eqref{eq:ecc_first_order} to find  the values of $\alpha$ in order to get the desired initial eccentricities $\epsilon_n$:
\begin{equation*}
    \epsilon_n = \frac{n\,\Gamma(n)}{ \Gamma(1+n/2)} \alpha R^{n-2}.
\end{equation*}
Notice that, apart from the case $n=2$, in which $\alpha$ is dimensionless, for every other $n$ its value depends on the initial transverse size:
\begin{gather*}
    \epsilon_2 \simeq 2\, \alpha;\\
    \epsilon_3 \simeq \frac{8\, \alpha\, R}{\sqrt{\pi}};\\
    \epsilon_4 \simeq 12\, \alpha\, R^2.
\end{gather*}
}
By doing so, we aim to study systems with fixed initial eccentricities which, as a consequence of the fireball evolution, develop the corresponding harmonic flows $v_n$.\\
In momentum space, following \cite{Romatschke:2003ms, Nopoush:2014pfa}, we use a more general version of the Romatschke-Strickland distribution function:
\begin{equation}\label{eq:modifiedRS}
    f(p) \propto \gamma_0\exp{ -\big[ \sqrt{p_x^2 (1+\psi_0) + p_y^2 + p_w^2 (1+\xi_0)}/\Lambda_0 \big]}
\end{equation}
Here $p_w$ is the momentum associated with $\eta_s$, defined as $p_w = p_z \cosh(\eta_s) - p_0 \sinh(\eta_s)$.\\
We determine $\gamma_0$ and $\Lambda_0$ in order to have the initial particle  and energy density associated to the given $T_0$ and $\Gamma_0$. These conditions correspond to determine the parameter $\Lambda_0$ and $\gamma_0$ by the following equations:
\begin{eqnarray}\label{eq:Lambdacondition}
    \Lambda_0 &=& T_0 \Bigg[ \frac{1}{2\sqrt{1+\xi_0}} + \frac{\sqrt{(1+\psi_0)(1+\xi_0)}}{4\pi} \times  \nonumber \\
    & & \int_0^{2\pi} d\phi \dfrac{ \arctan \sqrt{\frac{\xi_0 - \psi_0 \cos^2\phi}{1+\psi_0\cos^2\phi}} }{ \sqrt{(\xi_0 - \psi_0 \cos^2\phi)(1+\psi_0\cos^2\phi)^3 }  } \Bigg]^{-1} \\
    \gamma_0 &=& \frac{\Gamma_0 T_0^3}{\Lambda_0^3} \nonumber
\end{eqnarray}
We fix thorough the paper $T_0=0.5$ GeV and $\Gamma_0=1$. The two parameters $\xi_0$ and $\psi_0$ determine respectively the longitudinal and transverse anisotropy in momentum space. In particular, a $\psi_0\ne 0$ allows us to mimic an initial azimuthal anisotropy which corresponds to a finite $v_2(p_T)$, in order to simulate scenarios that in a Colour Glass Condensate approach could account for the measured finite $v_2$ in $pp$ and $pA$ \cite{Krasnitz:2002ng, Schenke:2015aqa, Lappi:2015vta,Mantysaari:2017cni, Greif:2017bnr, Schenke:2019pmk}.\\
We nearly always make us of the distribution function in Eq. \eqref{eq:modifiedRS}; however, in some specific case  we choose an initial distribution with $Y=\eta_s$, where $Y=\tanh(p_z/E)$ is the momentum rapidity. This corresponds to the limiting case of Eq. (\ref{eq:modifiedRS}) with  $\xi_0\to+\infty$ (or equivalently $p_w=0$) and $\psi_0=0$; Eqs. \eqref{eq:Lambdacondition} reduce to $\Lambda_0 = 3/2\, T_0$ and $\gamma_0=27/8\, \Gamma_0$.\\
\begin{figure}
    \centering
    \includegraphics[width=1\linewidth]{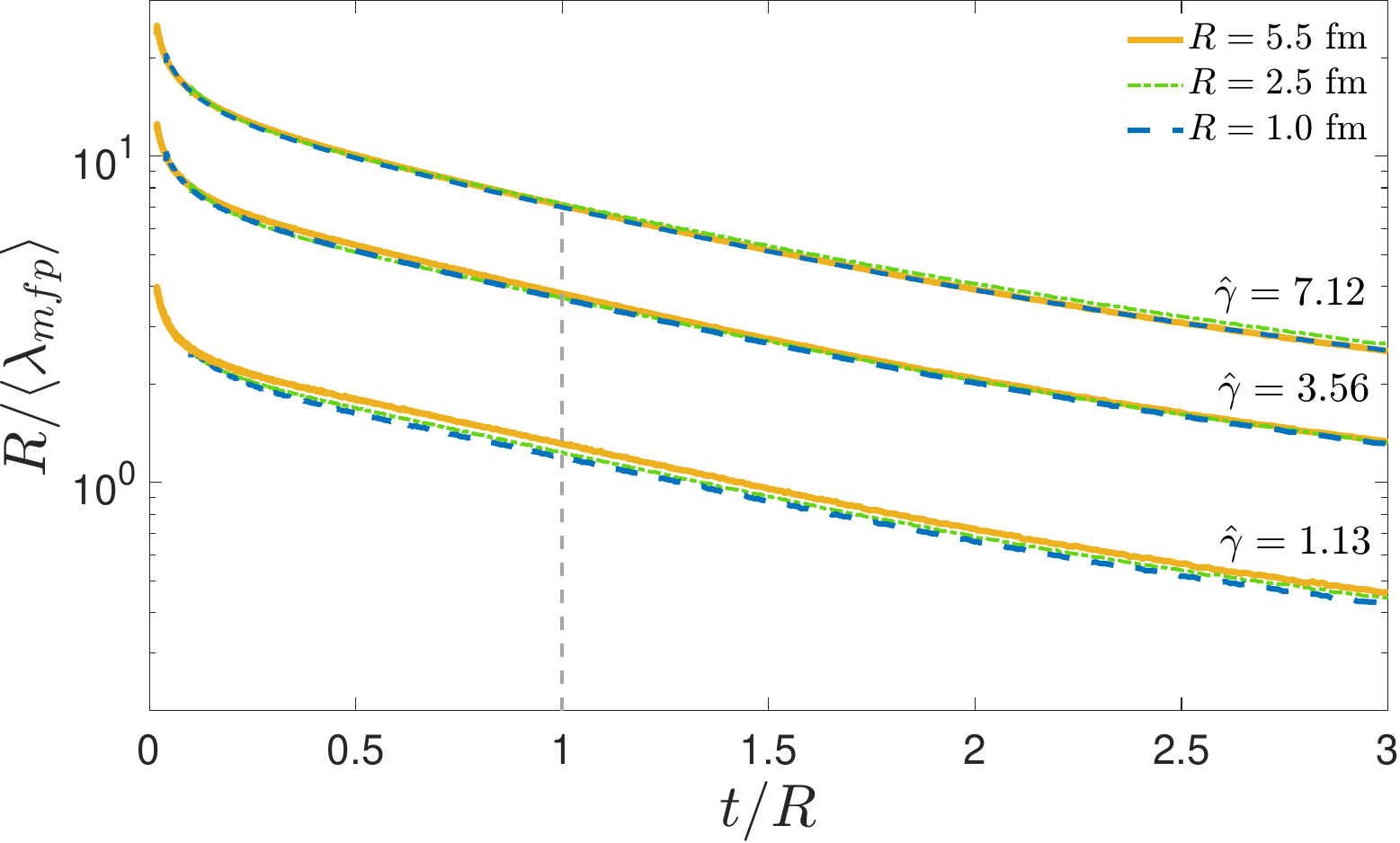}
        \caption{Ratio between transverse size $R$ and {average} mean free path $\langle\lambda_{\text{mfp}}\rangle$ as a function of scaled time $t/R$. Overlapped curves share the same opacity $\hat \gamma$. Corresponding $\eta/s$ values can be found in Table \ref{tb:opacity_values}.}
    \label{fig:gamma_def}
\end{figure}

\begin{table}[]
    \centering
    \begin{tabular}{ccc}
        \toprule
             {$\quad\hat \gamma\quad$} & {$\quad R\, \text{[fm]}\quad$} & {$\quad 4\pi\eta/s\quad $} \\
        \midrule
        \multirow{3}*{1.13} & 1.0 & 3.18  \\
            & 2.5 & 6.33  \\
            & 5.5 & 11.4  \\
        \midrule
        \multirow{3}*{3.56} & 1.0 & 1.00  \\
         & 2.5 & 2.00  \\
         & 5.5 & 3.61  \\   
        \midrule
        \multirow{3}*{7.12} & 1.0 & 0.503   \\
         & 2.5 & 1.00   \\ 
         & 5.5 & 1.81  \\
         \midrule
        \multirow{3}*{12.8} & 1.0 & 0.278    \\  
         & 2.5 & 0.554   \\ 
         & 5.5 & 1.00  \\
        \bottomrule
\end{tabular}
    \caption{Values of the opacity $\hat \gamma$ for different radii and shear viscosity over entropy density ratio. Initial temperature and proper time are fixed: $T_0=0.5$ GeV and $\tau_0=0.1$ fm.}
    \label{tb:opacity_values}
\end{table}

In 1D hydrodynamical and transport calculations \cite{Nugara:2023eku, Strickland:2018ayk, Strickland:2017kux} it is possible to identify a parameter which dominates the dynamical evolution and makes universal behaviour emerge disregarding microscopic details and different initial conditions. The cited works show how the evolution of the moments of the phase-space distribution function in terms of $\tau/\tau_{eq}$ depends solely on $(\tau/\tau_{eq})_0 \propto \tau_0 T_0 /(\eta/s)$, which contains information about the collision rate or, equivalently, the mean free path $\lambda_{\text{mfp}} \sim 1/n\sigma \sim T/(\eta/s)$, which, in the 1D case, is the only relevant physical scale. If we want to identify a similar quantity in a 3+1D simulation, we need to introduce a new parameter in which information is encoded also about the transverse size of the system. Previous works \cite{Kurkela:2018ygx, Kurkela:2018vqr, Kurkela:2019set, Ambrus:2021fej} have shown that such a parameter naturally emerges when rewriting the Boltzmann transport equation in RTA or ITA in a dimensionless fashion. It has been called opacity and defined as:
\begin{equation}\label{eq:opacity}
    \hat \gamma = \frac{1}{5\eta/s} \left( \frac{R}{\pi a} \frac{dE^0_\perp}{d\eta} \right)^{1/4} = \frac{ T_0}{5\,\eta/s} R^{3/4} \tau_0^{1/4},
\end{equation}
where $a=\varepsilon/T^4$ and in the RHS the opacity is expressed in terms of the quantities we fix.
As outlined in \cite{Kurkela:2018qeb} this parameter has the microscopic interpretation of $R/\lambda_{\text{mfp}} (0,R)$, i.e. the transverse system size $R$ in units of mean
free path $\lambda_{\text{mfp}}$ at $x_\perp=0$ and at the time $\tau=R$ at which the transverse expansion starts to dominate. In our approach, however, we are able to identify an equivalence class of different physical scenarios for each value of opacity $\hat \gamma$. If we consider the ratio between the two characteristic length scales of the system, the transverse size $R$ and the {mean free path $\langle\lambda_\text{mfp}\rangle$ averaged over the transverse plane}, we can verify that simulations sharing the same opacity show a {very similar} evolution of {$R/\langle\lambda_{\text{mfp}}\rangle$} in terms of $t/R$ at every time of the simulation, {with deviation $<5\%$,} as shown in Figure \ref{fig:gamma_def},
where we plot this quantity for three different values of $\hat \gamma = [1.13, 3.56, 7.18]$, each one for three different transverse radii $R=[1.0, 2.5, 5.5]$ fm. 
{This allows us to group different simulations in universality classes, depending on the $R/\langle\lambda_{\text{mfp}}\rangle$ behaviour;
each universality class can be labelled by the value of $R/\langle\lambda_{\text{mfp}}\rangle$ at $t=R$.
It should be emphasised that, at $t\approx R$, $R/\langle\lambda_{\text{mfp}}\rangle\approx \hat \gamma$ in each of the three cases. The slight difference could be traced back to the fact that we are integrating over the transverse plane (not only around $x_\perp= 0$) and that we do not make use of any estimation for the evolution of the involved quantities, but extract them at every time from the simulation.}
{We found the same behaviour considering also different initial profiles, in particular for a sharp circular distribution in the the transverse plane.}

\section{Attractors for the inverse Reynolds number}
\label{sec:reynolds}
\begin{figure}
    \centering
    \includegraphics[width=1\linewidth]{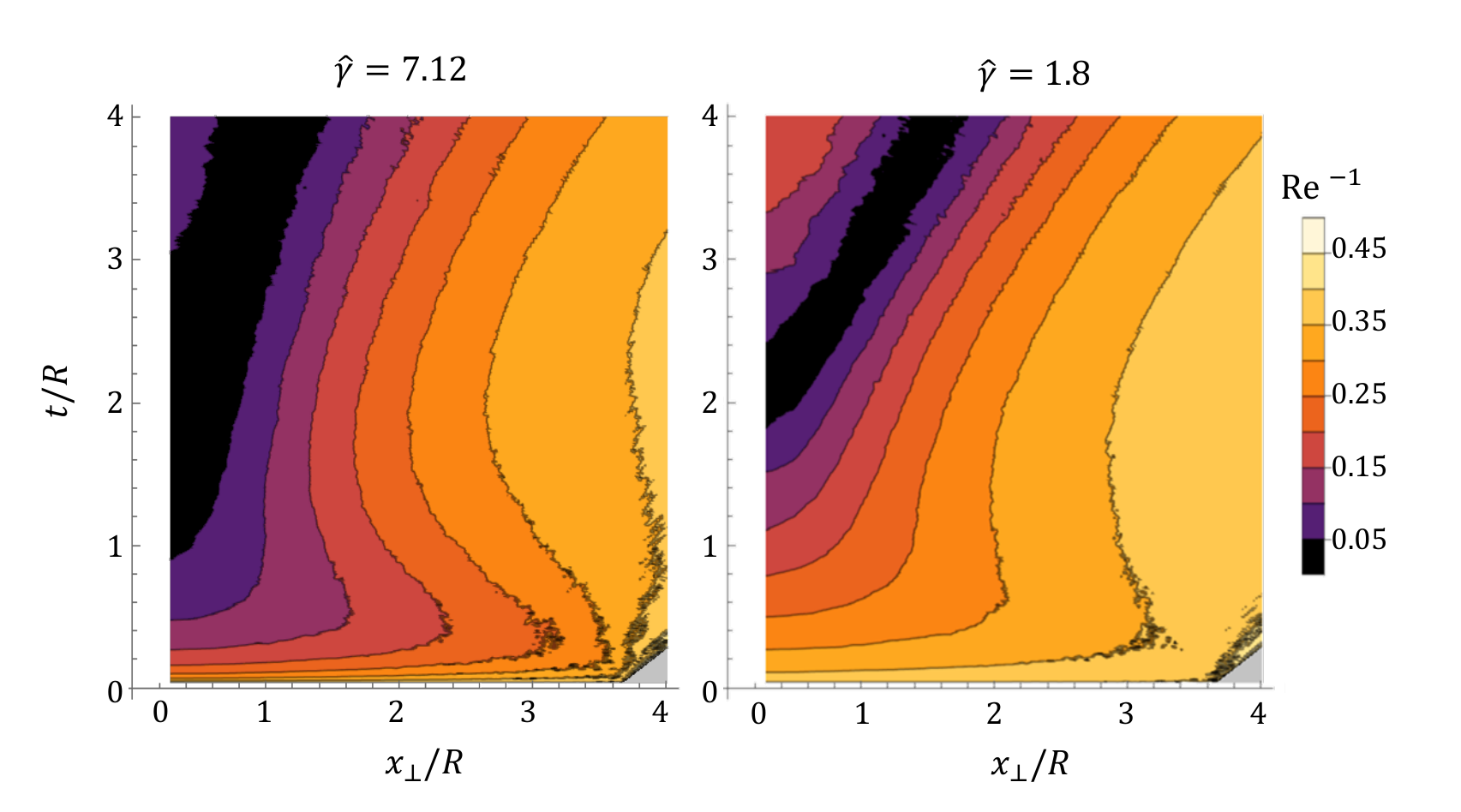}
    \caption{Contour plots of $\text{Re}^{-1}$ with respect to $t/R$ and $x_{\perp}/R$ at mid-rapidity $|\eta_s| < 0.125$ for two different values of opacity: $\hat\gamma = 7.12$ (left panel) and $\hat \gamma= 1.8$ (right panel). Both systems have $R=2.5$ fm; $\eta/s$ is respectively $1/4\pi$ and $4/4\pi$.}
    \label{fig:Q0_contour}
\end{figure}

\begin{figure}
    \centering
    \includegraphics[width=\linewidth]{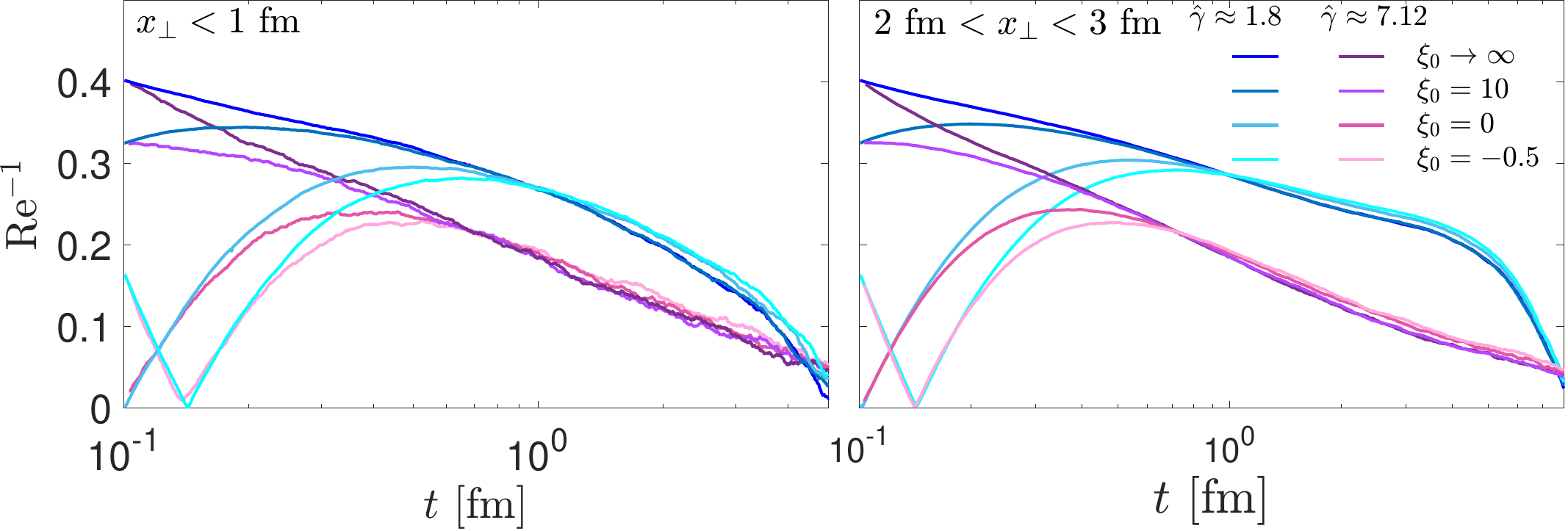}
     \caption{The Reynolds number as a function of $t$ for $\hat\gamma = [1.8, 7.12]$, respectively for $R=2.5$ fm ($\eta/s=4/4\pi$) and $R=5.5$  fm ($\eta/s=1.8/4\pi$) and for the two anuli $x_\perp<$ 1 fm (left panel) and 2 fm $<x_\perp<$ 3 fm (right panel); different shades correspond to different initial anisotropies.}
    \label{fig:reynolds_attractors}
\end{figure}

A macroscopic Lorentz scalar which quantifies the deviation of the fluid from the equilibrium is the inverse Reynolds number \cite{Denicol:book}, defined as:
\begin{equation}
    \text{Re}^{-1} = \dfrac{\sqrt{\pi^{\mu\nu}\pi_{\mu\nu}}}{\varepsilon} 
\end{equation}
where the shear stress tensor $\pi_{\mu\nu}$ in the conformal case is:
\begin{equation*}
    \pi^{\mu\nu} = T^{\mu\nu} - T^{\mu\nu}_{id}.
\end{equation*}
The ideal energy-momentum tensor is
\begin{equation*}
    T^{\mu\nu}_{id} = \text{diag}(\varepsilon, P, P, P)
\end{equation*}
and for the conformal case the pressure and the energy density are related by the equation of state $P(\varepsilon)=\varepsilon/3$. A vanishing inverse Reynolds number $\text{Re}^{-1}\to0$ corresponds to $\pi^{\mu\nu}\to 0$, which is characteristic of ideal hydrodynamics or of an equilibrated system. On the other hand, the free streaming evolution leads to an increase of the inverse Reynolds number.\\
Notice that, up to a numerical constant, the inverse Reynolds number is identical to the $Q_0$ parameter introduced in \cite{Kurkela:2019kip}:
\begin{equation*}
    Q_0= \sqrt{ \dfrac{\pi^{\mu\nu}\pi_{\mu\nu}}{T^{\mu\nu}_{id} T_{\mu\nu, id} } } = \sqrt{\dfrac{\pi^{\mu\nu}\pi_{\mu\nu}}{4/3\, \varepsilon^2}} = \frac{\sqrt{3}}{2} \text{Re}^{-1}.
\end{equation*}
In order to observe the evolution of systems with different values of opacity, we show two contour plots of $\text{Re}^{-1}$ with respect to $t/R$ and $x_{\perp}/R$ in Figure \ref{fig:Q0_contour}. 
Both plots refer to the same initial distribution, which is chosen to be isotropic in the transverse plane ($\psi_0=0$), with $Y=\eta_s$ and $R=2.5$ fm, but with different $\eta/s$, respectively $1/4\pi$ and $4/4\pi$. These values correspond respectively to $\hat \gamma = 7.12$ (left panel) and $\hat \gamma = 1.8$ (right panel). It is straightforward to observe that the larger the opacity, the wider the dark regions extend, thus showing that more interacting systems, i.e. with larger opacity values $\hat \gamma$, remain close to equilibrium in a wider spatial region and for longer times. It is also possible to see that at very small times and for large radii the system is dominated by the free streaming expansion, which causes a larger inverse Reynolds number. Qualitatively, these results are similar to what is shown in \cite{Kurkela:2019kip}.\\
It is interesting to study the $\invRe$ for different initial conditions and to look for universal behaviour in its evolution. In particular, we analyse systems with  different initial longitudinal anisotropy, which correspond to different initial $P_L/P_T$, for the two opacity values $\hat \gamma = [1.8, 7.12]$, corresponding respectively to $R=2.5$ fm and $\eta/s=4/4\pi$ and $R=5.5$ fm and $\eta/s=1.8/4\pi$. In order to do so, we fix $\psi_0=0$ and change $\xi_0=[-0.5, 0, 10, +\infty]$; since the system is azimuthally invariant, the inverse Reynolds number is a function of time and transverse radius $x_{\perp}$: $\text{Re}^{-1}=\text{Re}^{-1}(t,x_\perp)$. We consider two anuli $0<x_\perp<1$ fm and 2 fm $<x_\perp<$ 3 fm at midrapidity $|\eta_s|<0.125$. As said before, the inverse Reynolds number quantifies the deviation from equilibrium; as shown in Figure \ref{fig:reynolds_attractors}, after $t\approx 1$ fm, the curves approach a universal attractor regardless the initial conditions, therefore suggesting that, at this point, different systems reach the same degree of equilibration. Furthermore, we verified that the same evolution towards universality is reached also if a $\psi_0\ne 0$ is set in the initial distribution function, that corresponds to consider a non zero initial azimuthal anisotropy in momentum space. This leads to think to the possible presence of universal attractors also in {more differential probes of the phase-space distribution function, such as its momentum moments}.\\

\section{Attractors for Moments of $f(x,p)$}
\label{sec:moments}
In \cite{Nugara:2023eku} we studied the presence of universal attractors in the moments of the distribution function for conformal 0+1D and 1+1D systems using the RBT approach. We analysed the behaviour of the momentum moments of $f(x,p)$ defined as follows:
\begin{equation}
    M^{nm}(\tau) = \int dP (p\cdot u)^n (p\cdot z)^{2m} f(p,\tau),
\end{equation}
{with $u^\mu$ being the fluid 4-velocity and $z^\mu$ is a space-like vector orthogonal to $u^\mu$, which in the local rest frame of the fluid correspond to the $\hat z$ versor.}
In that case, being $f=f(p_w,p_T,\tau)$, this set of moments is a complete one.
Here, we extend 
this analysis to a more realistic case of 3+1D simulations studying systems with different transverse sizes, ranging from a typical fireball created in $AA$ collisions to smaller systems like $p$-Pb/$p$-Au and even $pp$ collisions at LHC and RHIC energies. For the results shown in this section we consider only azimuthally isotropic initial condition in coordinate and momentum space ($\epsilon_n=0$ and $\psi_0=0$), so that $f=f(x_\perp, p_T, p_w, \tau)$; this hypothesis will be relaxed in the next section.
The generalised moments are therefore:
\begin{equation}
    M^{nm}(x_\perp,t) = \int dP (p\cdot u)^n (p\cdot z)^{2m} f(x_\perp,p,t).
\end{equation}
It is useful also to define the integrated moments:
\begin{equation}\label{eq:moments_integrated}
    \mathcal M^{nm}(t) = \int_{A_\perp} d\mathbf{x_\perp} M^{nm}(x_\perp,t)
\end{equation}
where $A_\perp$ is the extension of the fireball in the transverse plane.
Following \cite{Strickland:2018ayk, Strickland:2019hff, Nugara:2023eku}, we look for attractors in the normalised moments $\overline M^{nm} = M^{nm}/M^{nm}_{eq}$, where

\begin{equation}\label{eq:moments_eq}
M^{nm}_{eq} (x_\perp, t) = \Gamma \int dP (p\cdot u)^n (p\cdot z)^{2m} e^{-(p\cdot u)/T},
\end{equation}

\begin{figure}
    \centering
    \includegraphics[width=0.5\linewidth]{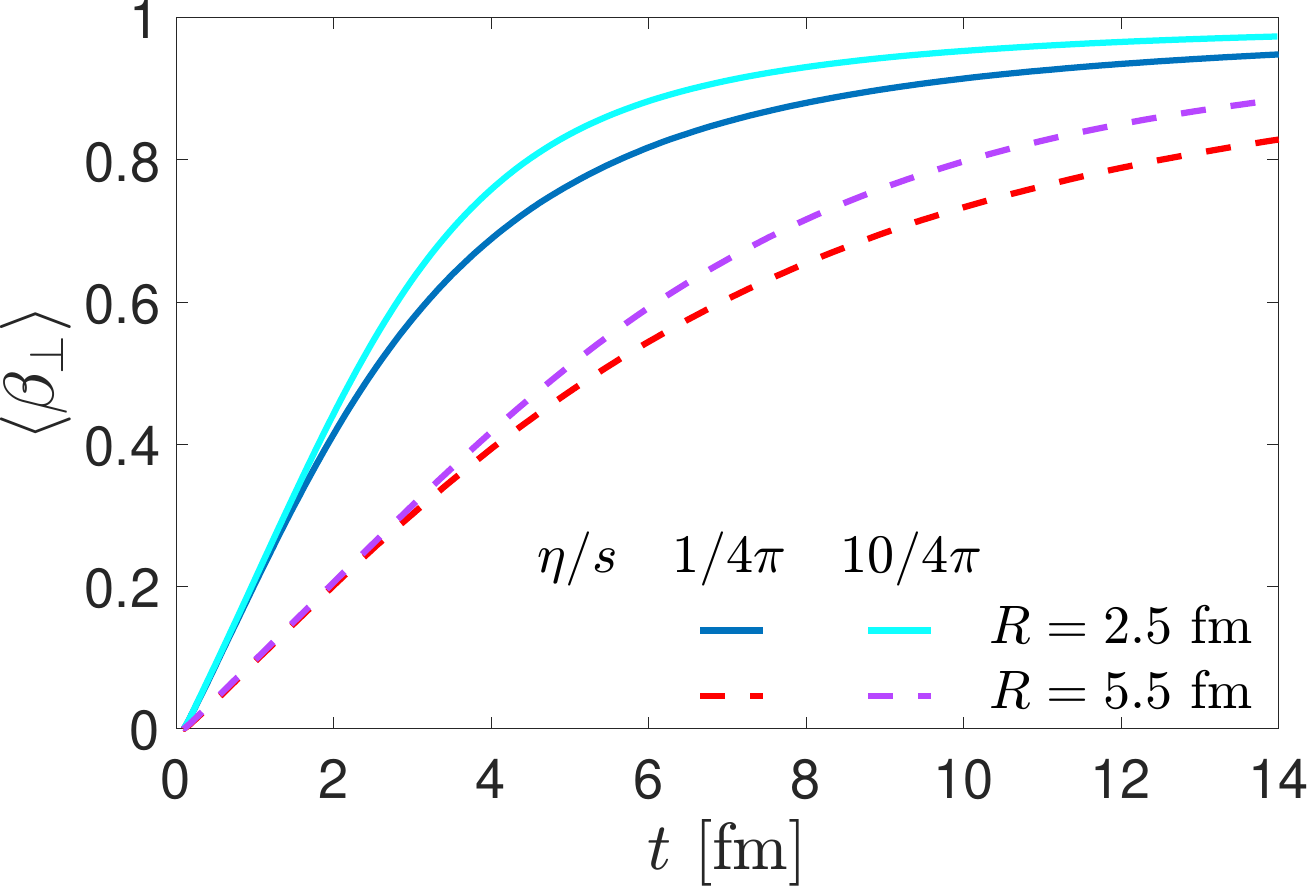}
    \includegraphics[width=0.439\linewidth]{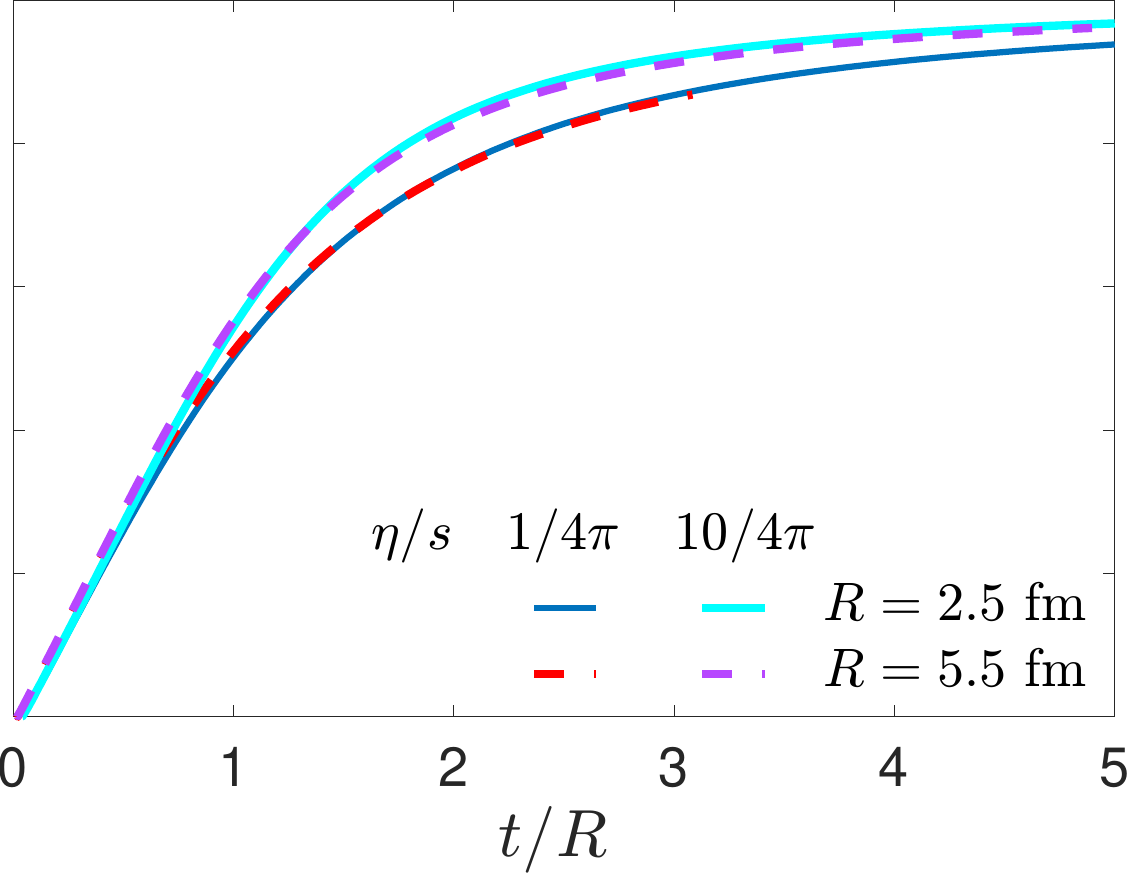}
    
    \caption{Average transverse flow $\langle \beta_\perp\rangle$  as a function of time $t$ (left panel) and scaled time $t/R$ (right panel). Simulations have same $T_0=0.5$ GeV, but different initial transverse size $R=2.5$ fm (solid lines), $R=5.5$ fm (dashed lines) and different specific viscosity $\eta/s=1/4\pi, 10/4\pi$. }
    \label{fig:beta}
\end{figure}
\begin{figure}
    \centering
    \includegraphics[width=0.85\linewidth]{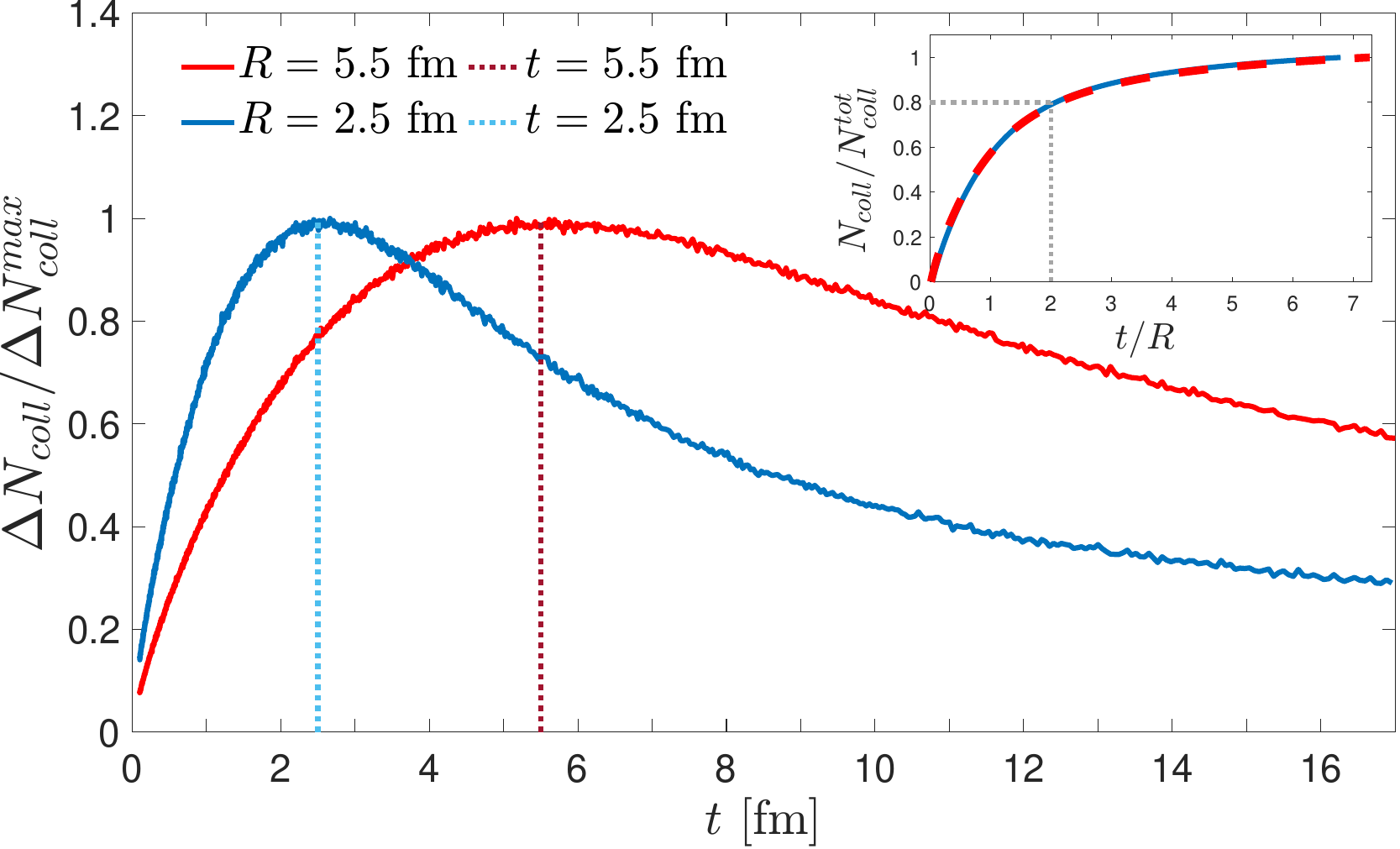}
    \caption{{Number of collisions in each time interval normalised by its maximum value and integrated over the transverse plane at mid-rapidity for two simulation with $\eta/s=1/4\pi$ and radii $5.5$ fm (red line) and $2.5$ fm (blue line). In the inset: the cumulative number of collisions as a function of $t/R$.}}
    \label{fig:collisions}
\end{figure}


where $\Gamma=\Gamma(x_\perp, t)$, $T=T(x_\perp, t)$, $u^\mu=u^\mu(x_\perp, t)$.\\
Before delving into the study of the moments, we must have a look to the onset of the flow $u^\mu$ which appears in the equilibrium distribution function in Eq. \eqref{eq:moments_eq}. Since the midrapidity region we are focusing on is symmetric in $\eta_s$ (and therefore in $z$), for symmetry reason it will be $\beta_z(x_\perp,t)=0$, as we checked out.  Therefore, in the rapidity region [-0.125, 0.125] the fluid velocity can be written as $u^\mu(x_\perp, t)=\gamma(1, \boldsymbol\beta_\perp(x_\perp, t), 0)$. In Fig. \ref{fig:beta} we show the average transverse flow $\langle\beta_\perp(t)\rangle$ in the midrapidity region for four different simulations with $R=[2.5, 5.5]$ fm and $\eta/s=[1/4\pi, 10/4\pi]$ . By looking at the left panel, it is interesting to notice that systems with different initial size $R$ and $\eta/s$ show a different evolution of the $\langle\beta_\perp(t)\rangle$. However, we observe that at $t \approx R$, $\langle \beta_\perp\rangle \approx 0.5$ as shown in the right panel of Fig.\ref{fig:beta}, which indicates also that, for $t>R$ the development of the flow is significantly non-zero. It is not surprising that curves corresponding to the higher value of $\eta/s$ show a faster development of the flow, since the system is closer to the free-streaming case.  {More generally, we observe that if we rescale the time with respect to the initial transverse radius $R$ the curves at fixed $\eta/s$ are almost indistinguishable: it must be emphasised that this scaling property does not depend on $\eta/s$ and is valid in a wide range of specific viscosity. It is interesting to notice that at $t\sim 2 R$, $\langle \beta_\perp\rangle \approx 0.8$, that is the system is very close to a free streaming regime, in which $\langle \beta_\perp\rangle \lesssim 1$.  This fact could be interpreted along with what seen in Fig. \ref{fig:collisions}, where we show the total number of collision in the transverse plane and at midrapidity for the two simulations with $\eta/s=1/4\pi$ and initial transverse radius $R=2.5$ fm (blue line) and $R=5.5$ fm (red line). Obviously, the same could be shown also for the case with $\eta/s=10/4\pi.$}
It is evident that at $t\sim R$ the system makes the maximum number of collisions and afterwards starts to decouple with fewer and fewer collisions. By looking at the inset, in which the cumulative number of collisions as a function of $t/R$ is shown, one can see that at $t\sim 2R$ both simulations have performed $\sim$80\% of the total number of collisions. At this time scale, indeed, the system is almost decoupled: the flat curve at large time indicates that no more collisions occur and thus the system is in a free-streaming regime. This is coherent with what we see in the average transverse flow $\langle \beta_\perp\rangle$. {In the time range $t=R- 2R$, in summary, the system turns from being dominated by the longitudinal expansion to an almost transverse free streaming, as previously observed in \cite{Ambrus:2022koq}}. Because of this, we choose as the final time of our simulations $t=2R-3 R$.\\
Numerically, our discretisation of the space-time allows us to calculate locally in each grid cell the parameters of Eq. \eqref{eq:moments_eq}, by making use of Eq. \eqref{eq:Landau_frame} and Eq. \eqref{eq:matching_conditions}. Therefore, Eq. \eqref{eq:moments_integrated} becomes:
\begin{eqnarray}
    \mathcal M^{nm}(t) = \sum_{\text{cell}}  M^{nm}(\mathbf x_{\perp}^{\text{cell}},t).
\end{eqnarray}
The same is done also for the equilibrium moments in Eq. \eqref{eq:moments_eq} where $\Gamma=\Gamma(x_\perp^\text{cell}, t)$, $T=T(x_\perp^\text{cell}, t)$, $u^\mu=u^\mu(x_\perp^\text{cell}, t)$. Once more the integrated moments are given by:
\begin{eqnarray}
    \mathcal M^{nm}_{eq}(t) = \sum_{\text{cell}}  M^{nm}_{eq}(\mathbf x_{\perp}^{\text{cell}},t).
\end{eqnarray}
{Notice that $\mathcal M^{10}=n$, $\mathcal M^{01}=P_L$ and $\mathcal M^{20}=\varepsilon$ and therefore, due to the matching conditions, $\overline {\mathcal M}^{10}=1$ and $\overline{\mathcal M}^{20}$=1, while $\overline {\mathcal M}^{01} = P_L/P_{eq} = 3P_L/\varepsilon$, which is strictly linked with $P_L /P_T = 2 P_L/(\varepsilon - P_L)$.}

\begin{figure}
    \centering
    \includegraphics[width=\linewidth]{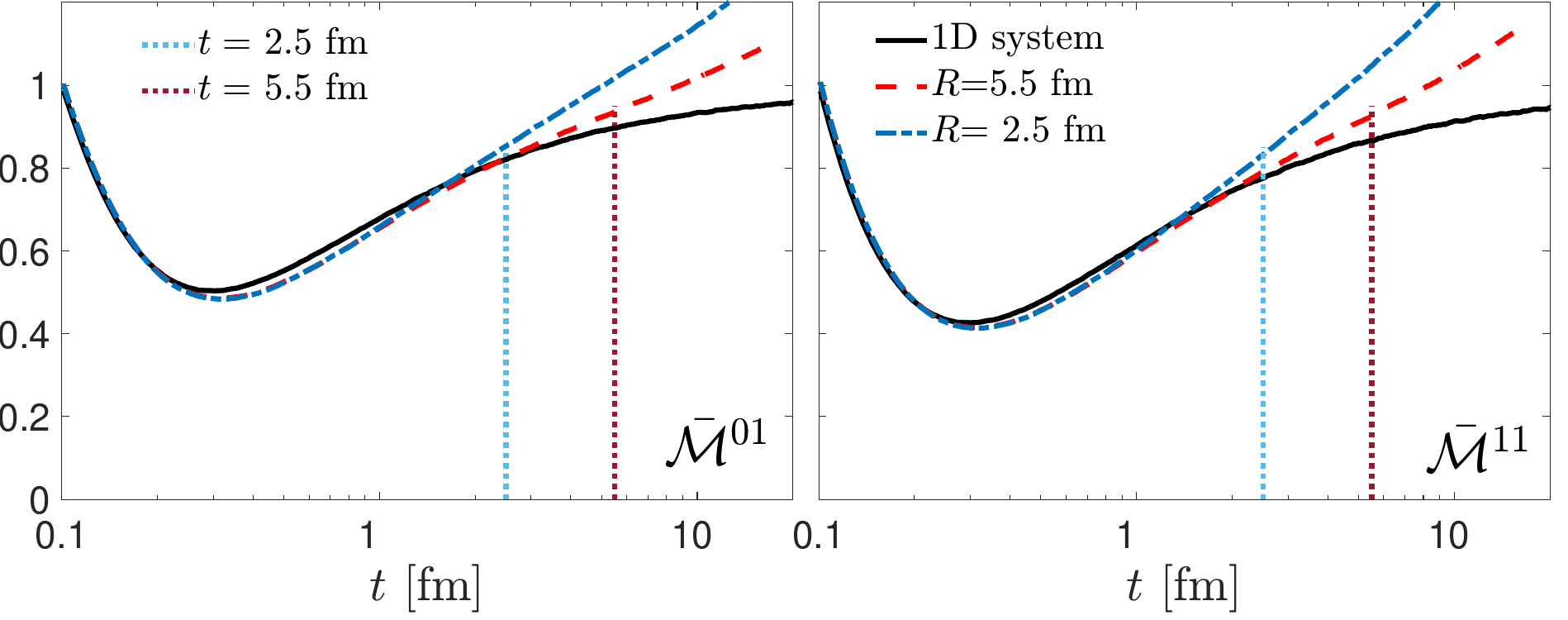}
    \caption{Normalised moments $\overline{ \mathcal M}^{01}$ and $\overline{ \mathcal M}^{11}$ as a function of time $t$. The 3D curves depart from the 1D one at a time scale $t\sim R$. All the simulations are performed with $T_0=0.5$ GeV, $\tau_0 =0.1$ fm, $\eta/s=1/4\pi$. It is possible to notice a slight dependence on the order of moments of the departure time, which is smaller for the larger order ones.}
    \label{fig:1d_vs_3d}
\end{figure}

In Fig. \ref{fig:1d_vs_3d} we show the integrated normalized moments $\overline {\mathcal M}^{nm}$ at midrapidity as a function of time. We compare the time evolution of the moments for two different system sizes $R=2.5$ fm and $R=5.5$ fm to the 1D case of \cite{Nugara:2023eku}. The three simulations share the same $\eta/s= 1/4\pi$ and same initial conditions: $T_0=0.5$ GeV, $\psi_0=0$, $\xi_0=0$. We observe that at very early times $t\ll R$ the three curves show the same evolution: as studied in 1D, there is a  departure from the initial equilibrium due to the longitudinal quasi-free streaming which dominates the early times, and a later increase of the normalised moments when the collision start to play a role. At time $t\approx R= 2.5$ fm the small system moments (blue dot-dashed line) depart from the 1D case; the same happens for the large ones at $t\approx R= 5.5$ fm (red dashed line). The deviation from the 1D case can be traced back to the development of the transverse flow $\beta_\perp$ observed in Fig. \ref{fig:beta}: as one can expect it is the transverse expansion which discriminate a 3D from a 1D simulation. Larger systems, which tend to develop later a transverse flow, behave for a longer time like a 1D system, which can be seen as a medium undergoing a longitudinal expansion with an infinite extension in the transverse plane. 

\subsection{Forward Attractors}
\begin{figure}
    \centering
    \includegraphics[width=.8\linewidth]{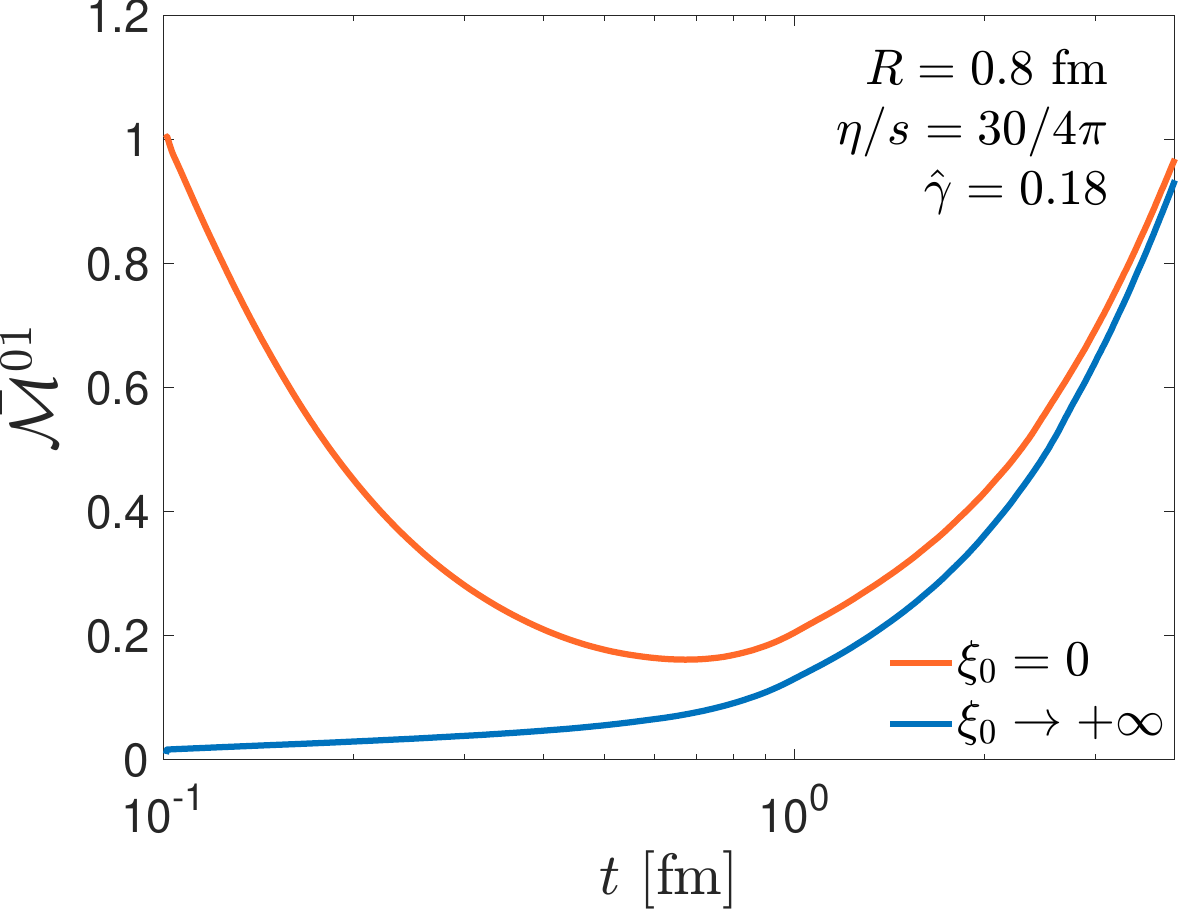}
    \caption{{Normalised integrated moment $\overline {\mathcal M}^{01} = P_L /P_{eq}$ for a simulation with very small opacity $\hat \gamma = 0.18$. Notice that there is no attractor even going up to $t=4$ fm, i.e. $t=5R$.}}
    \label{fig:moments_noattractor}
\end{figure}

In this section we study the normalised moments evolution for the same system sizes discussed above and $\eta/s=1/4\pi$, but for different initial anisotropy values $\xi_0=[-0.5, 0, 10, \infty]$ in order to highlight the emergence of universal behaviour.
\begin{figure}
    \centering
    \includegraphics[width=\linewidth]{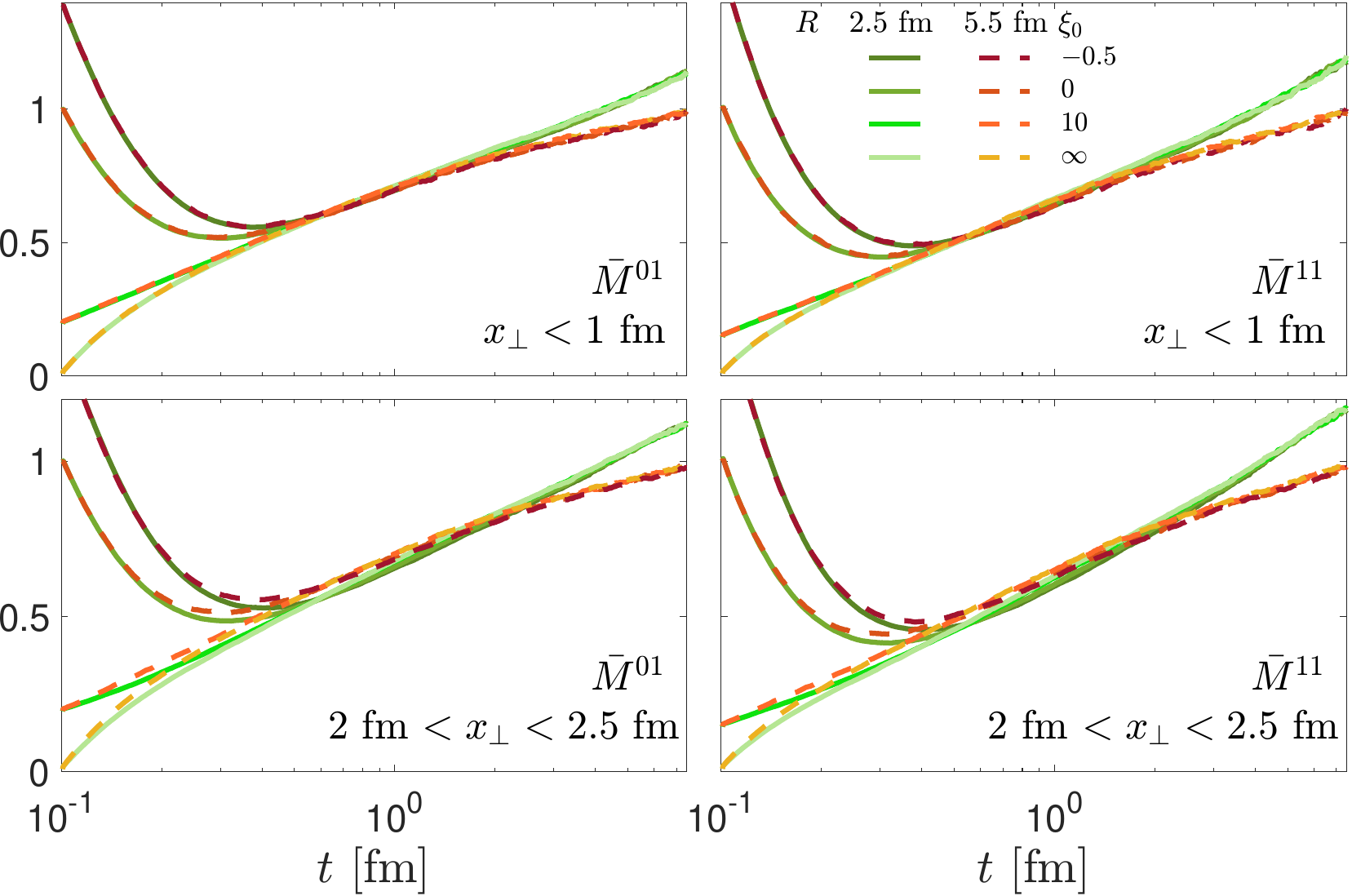}
    \caption{Forward attractors as a function of time in the two moments $\overline M^{01}, \overline M^{11}$ of the distribution function, for two different anuli: $x_\perp<1$ fm (upper panels) and 2.0 fm $<x_\perp<$ 2.5 fm (lower panels) and two different initial transverse sizes $R=2.5$ fm (green solid lines) and $R=5.5$ fm (orange dashed lines). Both computations are performed with $\eta/s=1/4\pi$. }
    \label{fig:forward}
\end{figure}
In Fig. \ref{fig:forward}, we show the normalised moments $\overline M^{nm} (x_\perp, t)$ for two different anuli $x_\perp<1$ fm (upper panels) and 2.0 fm $<x_\perp< 2.5$ fm (lower panels). We observe that, irrespectively of the initial size $R$, the systems reach the attractor at the same time scale, thus
forgetting about the different initial anisotropy values immediately after the minimum in the curves. As studied in \cite{Nugara:2023eku, Heller:2020anv, Jankowski:2023fdz}, this universal behaviour is due to the strong initial longitudinal expansion of the system, which makes it lose memory about the initial conditions; since, as seen in Fig. \ref{fig:1d_vs_3d}, the initial expansion in these cases is identical to that of a 1D system, it is clear that the same loss of memory has to occur also in these 3D systems.\footnote{Notice that the universality here shown for different anuli is obviously present also in the integrated normalised moments $\overline {\mathcal M}^{nm}$.}\\
However, at later times the curves depart from the typical 1D trend and therefore the attractor sensitively depends on the initial system size $R$.\\
Nonetheless, one can imagine a more extreme scenario in which the transverse expansion takes place before the longitudinal one has been completed. Obviously this is a combined effect of the interaction strength of the system, encoded in the specific viscosity $\eta/s$, which determines how long the longitudinal expansion lasts, and of the initial transverse radius $R$, that sets the time scale at which the transverse flow sets off. By accurately choosing such quantities, one can get a system which keeps memory of the initial anisotropy, with the consequent lost of the universal behaviour, {as shown in Fig. \ref{fig:moments_noattractor}. It is possible to see that the attractor is not reached even for $t=4=5R$ fm, 10 times larger than $t=0.4$ fm, at which the forward attractor is reached in Fig. \ref{fig:forward}. For this simulation we choose a very small radius $R=0.8$ fm and $\eta/s=30/4\pi$ (the system is very weakly interacting), in order to have $\hat \gamma = 0.18$. Notice that such an extreme case is far from possible systems created in uRHICs; nonetheless it quite instructive in order to understand the interplay of the two time scales, i.e. the time needed for the longitudinal expansion and the transverse size $R$, which marks the onset of the transverse expansion.}

\subsection{Pullback Attractors}

\begin{figure}
    \centering
    \includegraphics[width=\linewidth]{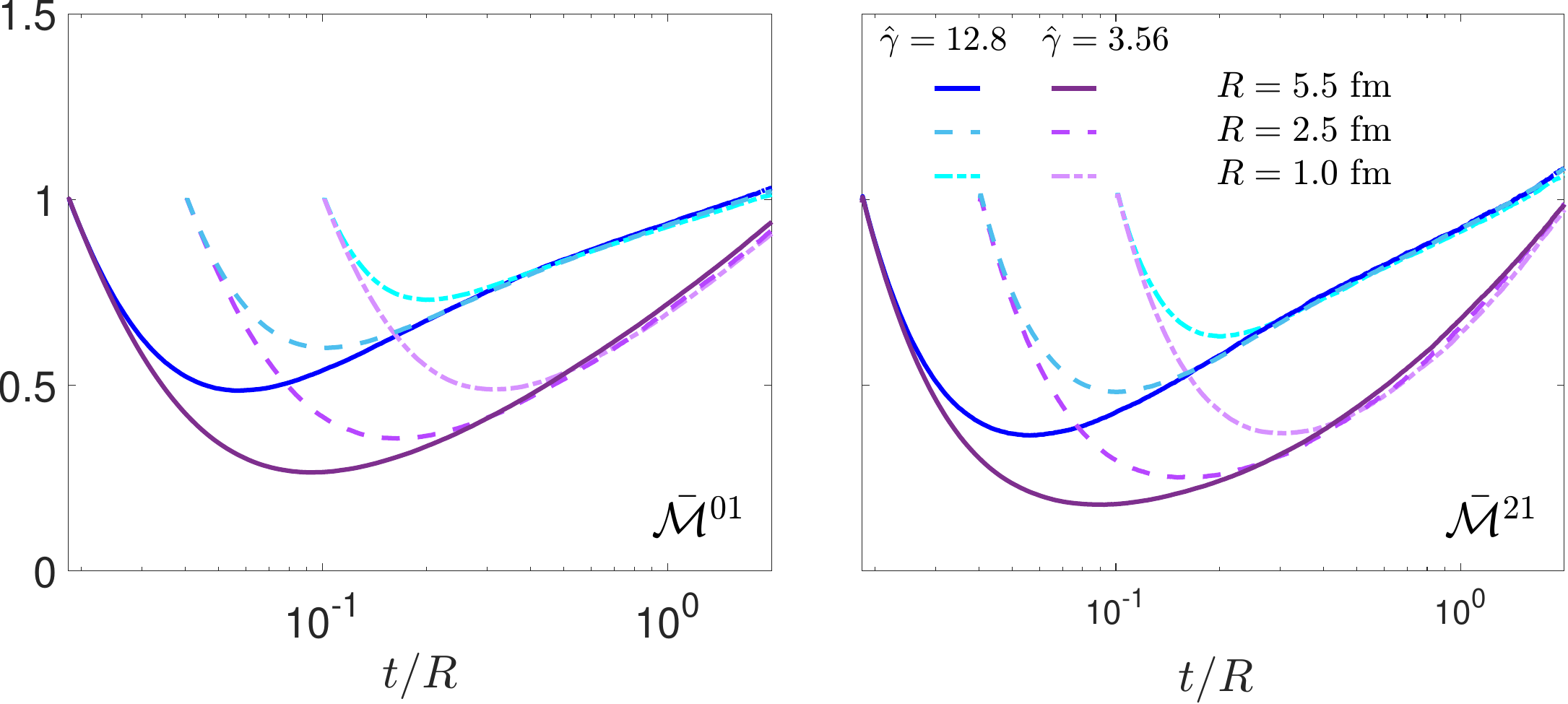}
    \caption{Comparison of two different pullback attractors for different values of the opacity $\gamma$. Notice that the equilibrium line is crossed at very similar scaled times.}
    \label{fig:double_pullback}
\end{figure}

In  0+1D hydrodynamics, relativistic kinetic theory calculations \cite{Strickland:2017kux, Alalawi:2020zbx,Kurkela:2018vqr, Kurkela:2019set,Strickland:2018ayk, Heller:2018qvh,Blaizot:2017ucy} and 1+1D transport approach \cite{Nugara:2023eku} it has been observed that universal behaviour of the normalised moments plotted with respect to the scaled $\tau/\tau_{eq}$ is obtained when $(\tau/\tau_{eq})_0\sim \tau_0 T_0 /(\eta/s) $ is changed. This pattern is called pull-back attractor.  This is directly linked with the typical initial expansion of 0+1D or 1+1D systems and to the fact that $\tau_{eq}$ is the only time scale dominating the evolution. However, in the 3+1D case, as mentioned above, there is a second time-scale strongly characterising the system, which is related to the expansion in the transverse direction and can be specified by the transverse size $R$. As anticipated before, one can make use of the dimensionless parameter $\hat \gamma = T_0 R^{3/4} \tau_0^{1/4} / (5\eta/s)\propto R/\lambda_{\text{mfp}}$ which includes also information about $R$, see Eq. \eqref{eq:opacity}. It may be of interest to rewrite the opacity as:
$$\hat \gamma = \frac{\tau_0 T_0}{5 \eta/s} \left( \frac{R}{\tau_0} \right)^{3/4} = (\tau/\tau_{eq})_0 \left( \frac{R}{\tau_0} \right)^{3/4}.$$
In such a way, the new parameter $\hat \gamma$ is directly related to the ratio $(\tau/\tau_{eq})_0$, which characterises the 1D systems, while the factor $(R/\tau_0)^{3/4}$ accounts for the transverse extension. It is easily to argue that, differently from the 1D case, we cannot expect our simulations to exhibit universality by fixing $\tau_0 T_0/(\eta/s)$, as will be clear below.
\\
In this subsection, we want to investigate if the universal behaviour of the pull-back attractors, which is qualitatively  different from what studied in the previous subsection,  emerges also in 3+1D simulations.
In Fig. \ref{fig:double_pullback} we show the integrated normalised moments as a function of $t/R$ for simulations corresponding to several system size values $R$ and viscosity $\eta/s$ ranging from typical $pp$ to $AA$ collisions. It is straightforward to see that the three violet curves converge to a unique behaviour and the same is true for the three blue curves, in both cases for $t/R<1$: these are the equivalent of the pull-back attractors observed in the 1D case. The two groups are characterised by two different values of opacity $\hat \gamma$, respectively $\hat \gamma = 3.56$ and $\hat \gamma = 12.8$, and the same behaviour has been observed for a wider range of $\hat \gamma$, down to $\hat \gamma \lesssim 1$. Therefore the opacity defines a universality class in which systems share the same behaviour in terms of normalised moments after a certain scaled time $t/R$ which always occurs immediately after the minimum, that is when the collisions start to dominate with respect to the initial longitudinal expansion. To our knowledge, this is the first time such a pattern is highlighted in 3D+1 systems. 

\begin{figure*}[t]
    \centering
    \includegraphics[width=.32\linewidth]{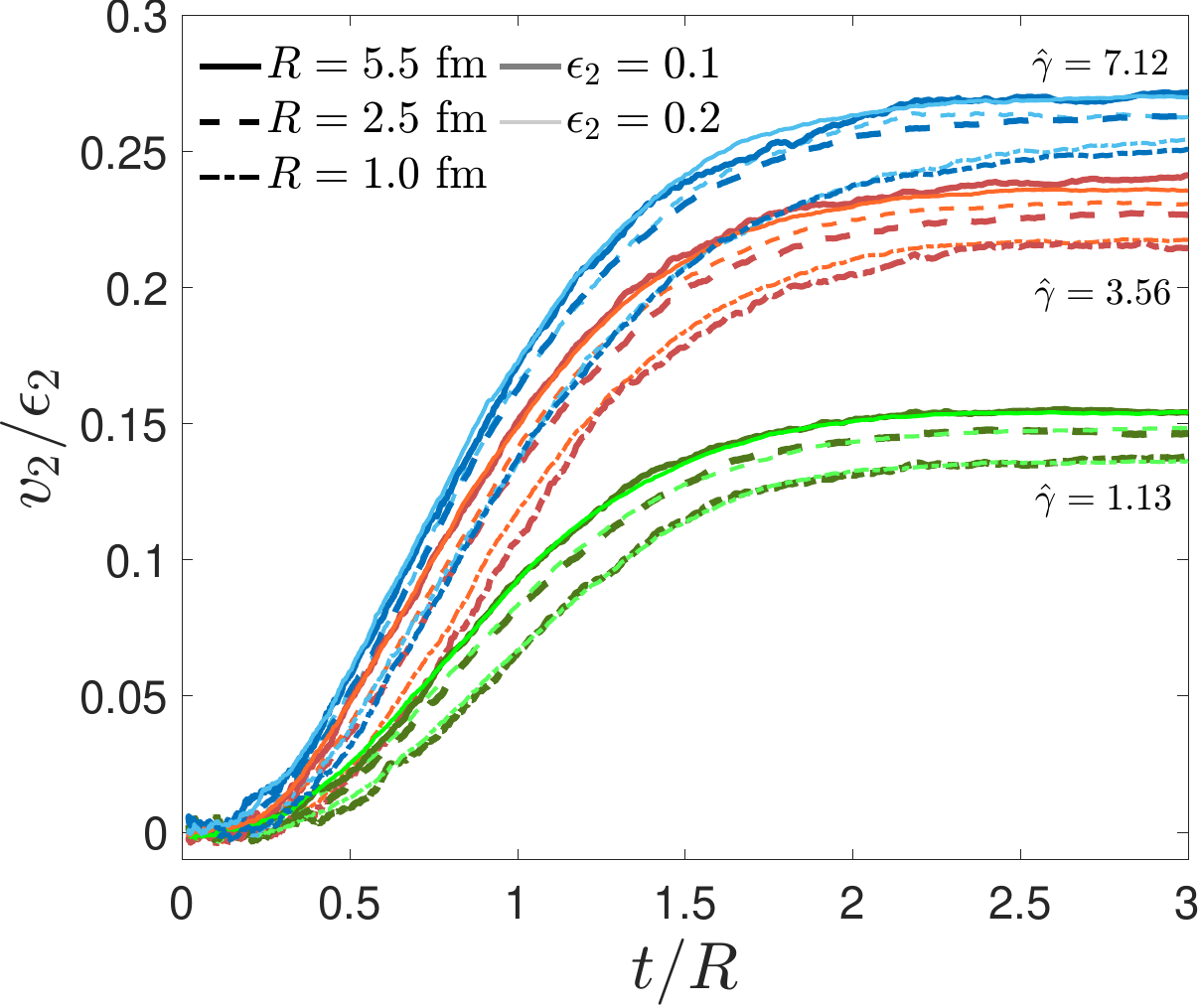}
    \includegraphics[width=.32\linewidth]{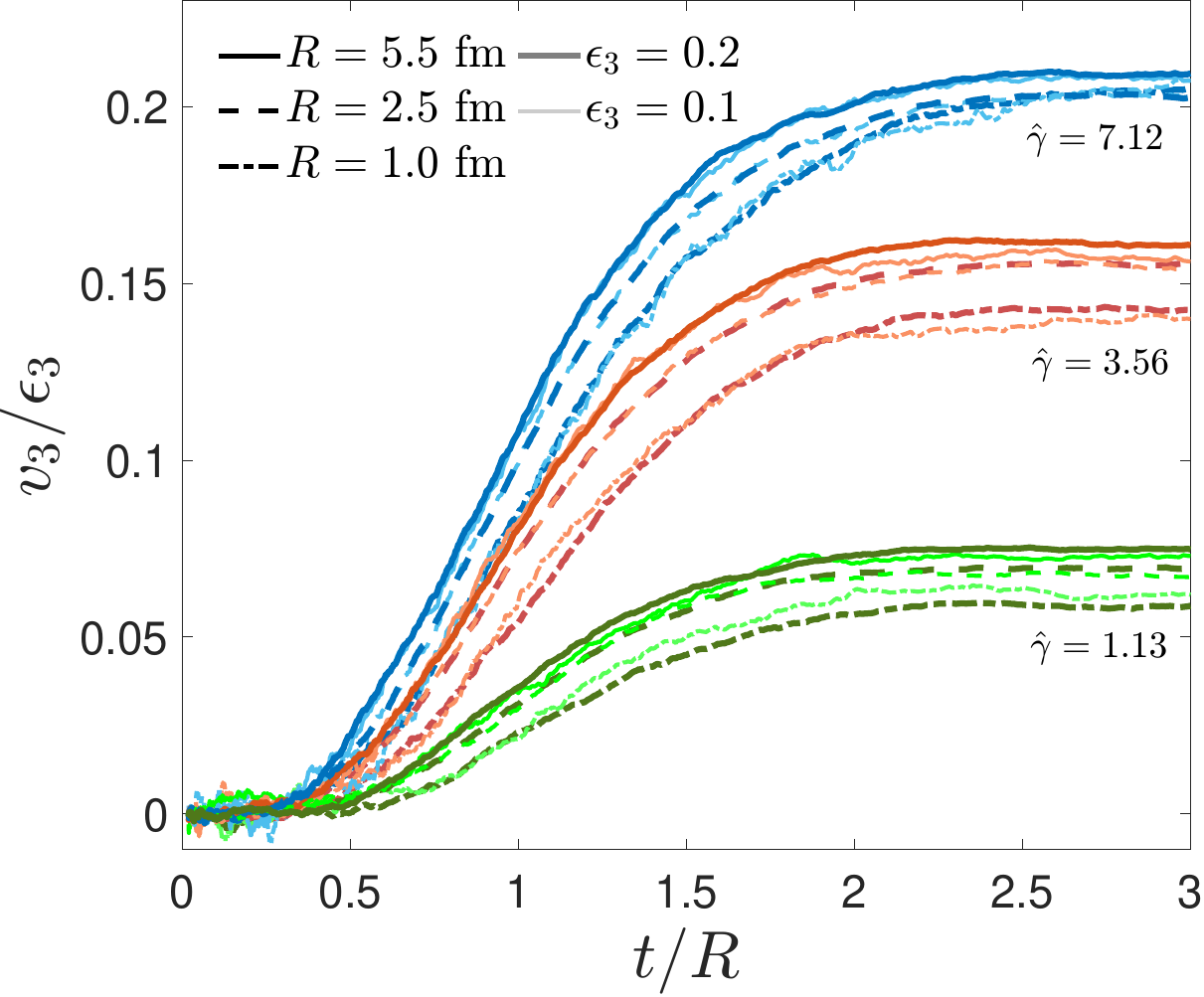}
    \includegraphics[width=.32\linewidth]{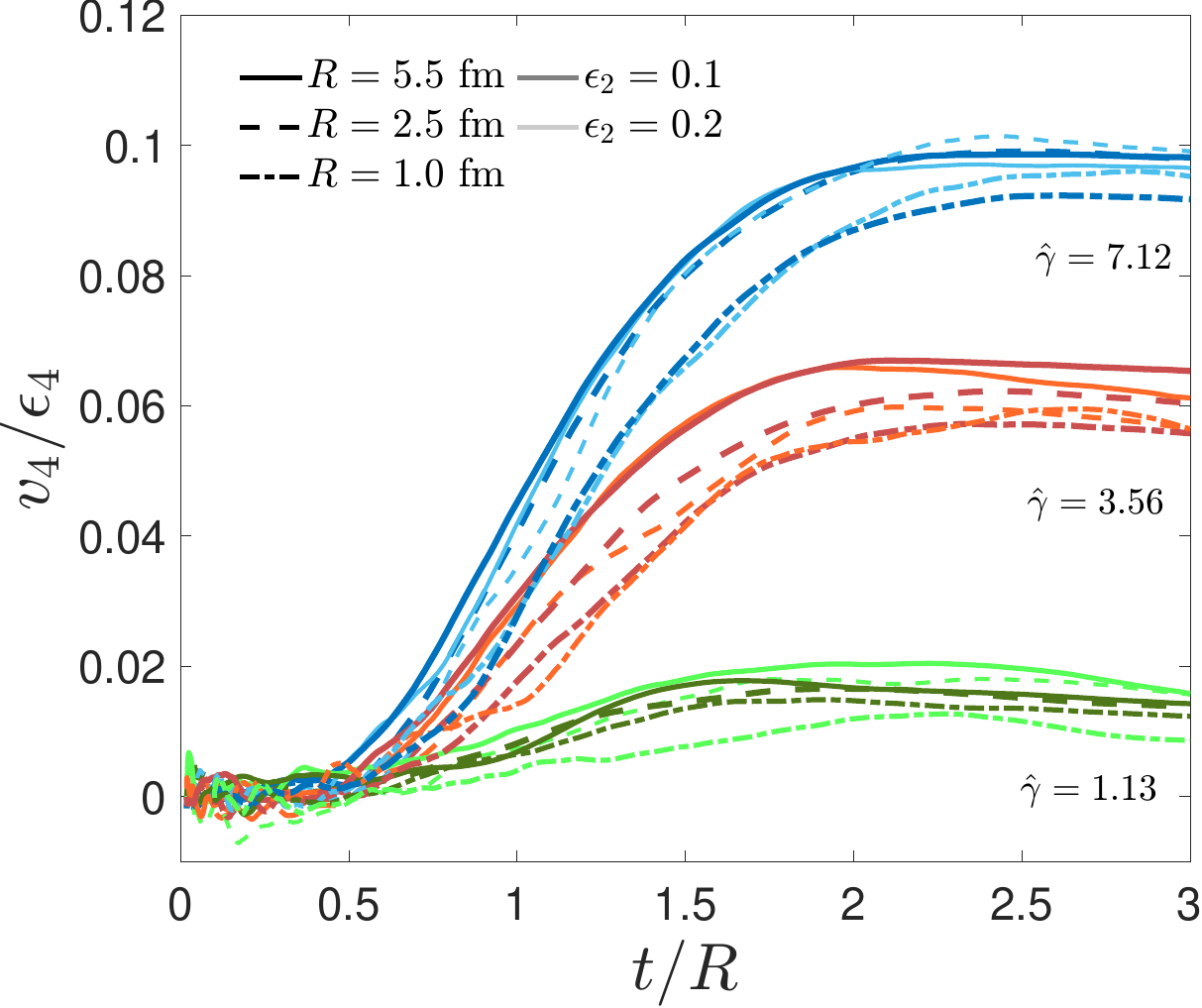}
    \caption{Response function $v_n/\epsilon_n$ for different values of initial eccentricities and different values of opacity $\hat \gamma$. The details on the corresponding $\eta/s$ can be found in Table \ref{tb:opacity_values}.}
    \label{fig:response_vn}
\end{figure*}

\section{Study of the anisotropic flows}
\label{sec:flows}
In this section we analyse the time evolution of the anisotropic flow coefficients $v_n$ for different values of transverse size $R$ and specific viscosity $\eta/s$, in order to explore a wide range of opacity values $\hat \gamma$, going from small systems like $pp$ or $pA$ to larger ones ($AA$). In our approach we generate a final anisotropic flow $v_n$ by implementing an initial eccentricity according to Eq. \eqref{eq:eccentricity}. We choose the $\alpha$ values for different $n$ in order to get $\epsilon_n=[0.1, 0.2]$ which correspond to typical hypothesised initial eccentricities produced in central and mid-peripheral collisions \cite{Plumari:2015cfa}.\\

The anisotropic flow coefficients allow us to characterise the particle distribution function, which indeed can be expanded as:
\begin{equation*}
    \frac{dN}{d\phi\,p_\perp\,dp_\perp} \propto 1 + 2 \sum_{n=1} v_n (p_\perp) \cos[ n(\phi_p - \Psi_n(p_\perp)) ].
\end{equation*}
In our approach there is no $n$-dependent event plane, therefore we choose $\Psi_n=0$. In particular, we are interested in the $v_n$ coefficients which can be computed as:
\begin{equation*}
    v_n (p_\perp) =  \frac{\displaystyle\int d^2\mathbf x_\perp \displaystyle\int \dfrac{dp_z}{2\pi} \displaystyle\int d\phi_p\,  \cos ( n\phi ) \,f(x,p)}{\displaystyle\int d^2\mathbf x_\perp \displaystyle\int \dfrac{dp_z}{2\pi} \displaystyle\int d\phi_p\, f(x,p)}.
\end{equation*}
We will make use of the integrated coefficients: in the formula above we integrate, both in the numerator and in the denominator, not only over $p_z$ and $\phi_p$ but over the three momentum $d^3\mathbf p$:
{
\begin{equation}\label{eq:intergrated_vn}
         v_n= \frac{\displaystyle\int d^2\mathbf x_\perp \displaystyle\int \dfrac{d^3\mathbf p}{(2\pi)^3}  \cos ( n\phi ) f(x,p)}{\displaystyle\int d^2\mathbf x_\perp \displaystyle\int \dfrac{d^3\mathbf p}{(2\pi)^3} f(x,p)}.
\end{equation}
}

\subsection{Time evolution of the response functions $v_n/\epsilon_n$}

We study the time evolution of the linear response ratio $v_n/\epsilon_n$ for $n=[2,3,4]$. In Fig. \ref{fig:response_vn}
we show the response functions as a function $t/R$ for three different values of $\hat \gamma$ = [1.13, 3.56, 7.12]. For fixed $\hat \gamma$ we explore three different initial sizes $R=[1.0, \, 2.5, \, 5.5]$ fm so that 
the value of $\eta/s$ is determined by the definition of $\hat \gamma$ (see Table \ref{tb:opacity_values}). Thinner, lighter lines correspond to $\epsilon_n=0.1$, while bolder, darker lines to $\epsilon_n=0.2$.\\
We observe that the curves cluster in three branches according to the opacity $\hat \gamma$, thus agreeing, in a first approximation, with the approach of other studies within RTA and ITA such as \cite{Ambrus:2022oji, Kurkela:2019kip, Kurkela:2020wwb}. As expected, the larger the value of the opacity, the more efficiently the system is able to convert the initial anisotropy in coordinate space to the momentum space.
Going into detail, however, we notice a difference in the response function up to 10\% within the same group with fixed $\hat \gamma$, with a monotonic behaviour with the initial transverse size $R$.     Since $\hat \gamma \propto R^{3/4}/(\eta/s)$, keeping fixed $\hat \gamma$ and moving to higher $R$ means also going to higher $\eta/s$. \\
We also observe that the third harmonic flow $v_3$, despite showing a similar behaviour, exhibit a larger separation between different opacity branches.\\
\begin{figure}[h]
    \centering
    \includegraphics[width=.49\linewidth]{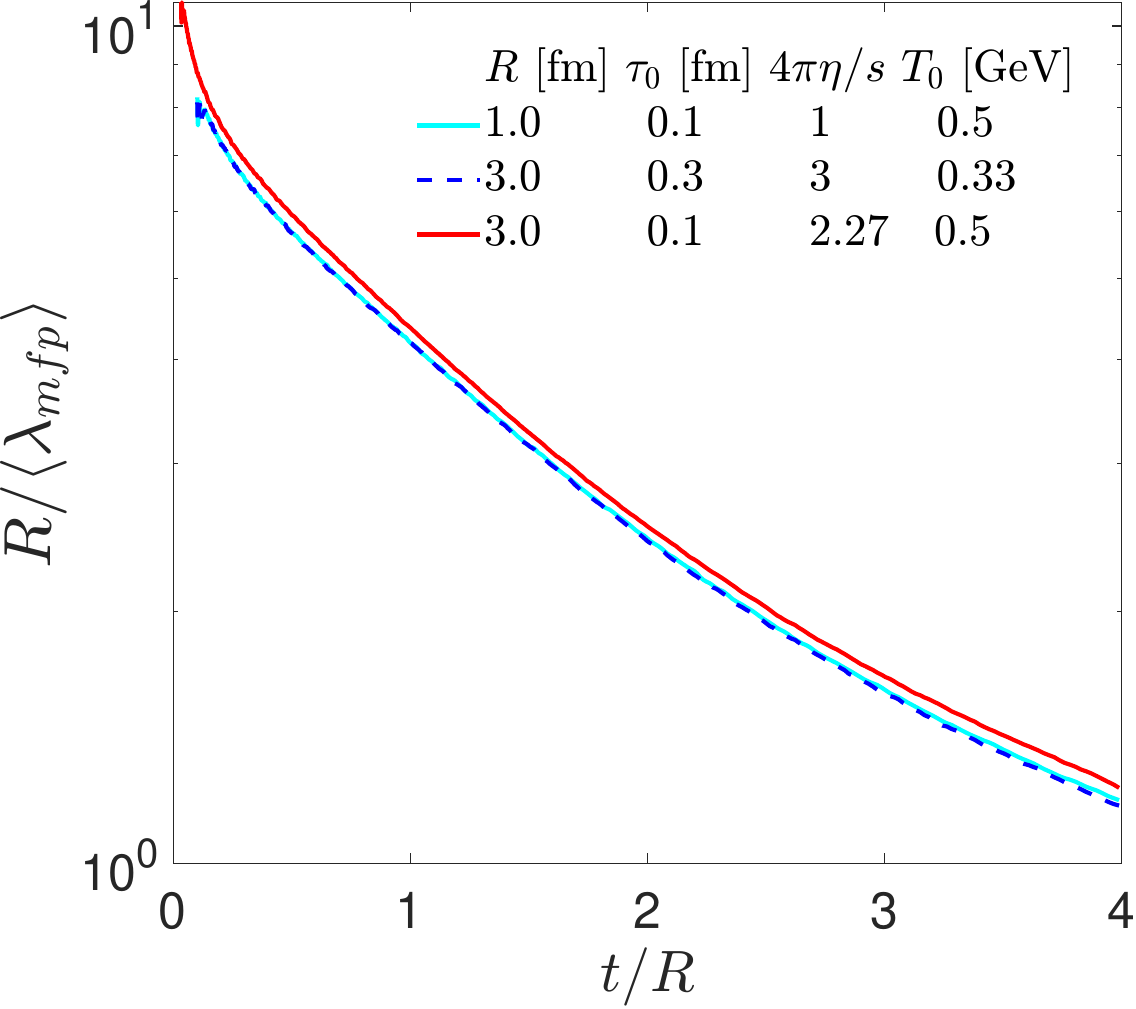}
    \includegraphics[width=.49 \linewidth]{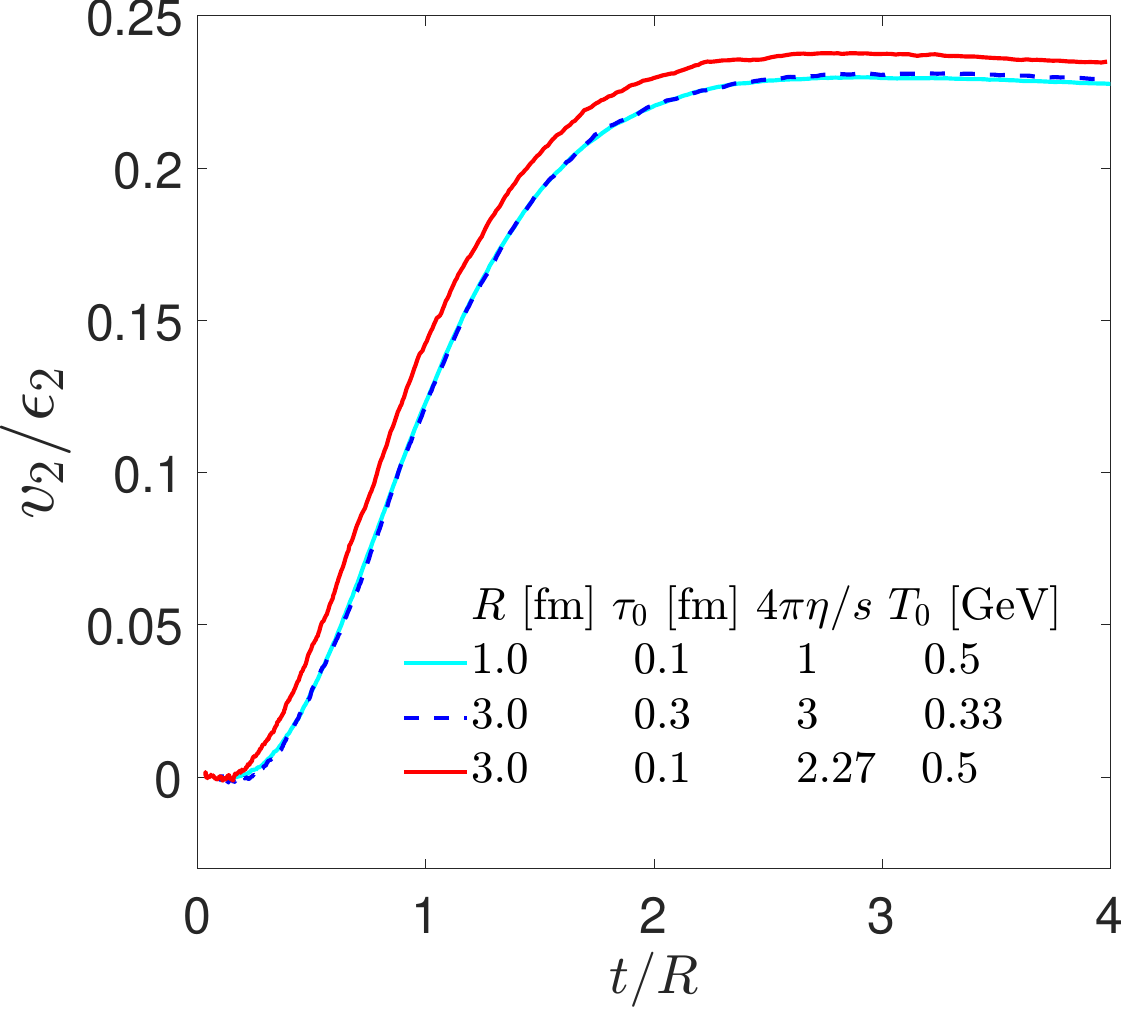}
    
    \caption{On the left panel we show $R/\langle \lambda_{\text{mfp}}\rangle$ as a function of $t/R$ for three different cases. For the solid cyan and dashed blue curves the parameters have been chosen in order to have exactly the same $R/\langle \lambda_{\text{mfp}}\rangle$, while the solid red curve, despite sharing the same $\hat \gamma=3.56$ of the other two, is slightly different. On the right panel we show the $v_2/\varepsilon_2$ for the same simulations: as one can see the cases with the same $R/\langle \lambda_{\text{mfp}}\rangle$ show an identical time evolution of the response curve.  }
    \label{fig:exactlyRlambda}
\end{figure}
{However, as one can see by looking in detail at Fig.\ref{fig:gamma_def}, different curves in the same opacity class show a slight deviation in $R/\langle \lambda_{\text{mfp}}\rangle$, which can be held responsible for the discrepancies observed in the $v_n/\varepsilon_n$. Moreover, we checked that the deviation in the response functions increase with the increasing of the difference in the $R/\langle \lambda_{\text{mfp}}\rangle$ ratio. We also performed calculations in which we kept exactly fixed $R/\langle \lambda_{\text{mfp}}\rangle$ for different values of $R$, $\eta/s$, $T_0$ and $\tau_0$, finding the same time evolution of the response functions in the range of parameters explored, as showed for a specific case in Fig.\ref{fig:exactlyRlambda}. This suggests that $R/\langle \lambda_{\text{mfp}}\rangle$ may encode the most relevant information about the underlying microscopic dynamics. However, in this paper we are developing the whole study keeping fixed $\hat \gamma$ with the aim to compare with previous literature in RTA and ITA approximation; we deserve for future works to develop a similar study keeping fixed $R/\langle \lambda_{\text{mfp}}\rangle$.}\\
Notice that the $v_n/\epsilon_n$ curves all saturate at $t\sim 2R$, that is what we expect since at this time scale the system is about to decouple (see the inset in Fig. \ref{fig:collisions}). Moreover we observe that the curves are independent on the initial eccentricity $\epsilon_n$, at least in the regime of small deformations.
Even though we are mainly interested in the behaviour of the response function at $t\to \infty$, we also looked at early times $t<1$ fm to analyse how the anisotropic flows are developed. We checked that, in agreement with previous studies \cite{Borghini:2022qha, Borrell:2021cmh}, in the particle-like regime (small $\hat \gamma$) $v_n\propto t^{n+1}$, while in the hydrodynamic regime (large $\hat \gamma$)  $v_n\propto t^{n}$, with the exponent smoothly going from $n$ to $n+1$ with increasing $\hat \gamma$. In addition, we found that, for a fixed $\hat \gamma$ value, a similar dependence is present also on $R$, with larger values of $R$ providing smaller exponents. As far as the $v_2$ is concerned, we found, within the investigated range of $R=1-5.5$ fm, $n=1.9-2.3$ for $\hat \gamma = 7.12$, while $n=2.4-2.9$ for $\hat \gamma=1.13$. A similar trend is observed also for $v_3$ and $v_4$.

\begin{figure}
    \centering
    \includegraphics[width=1\linewidth]{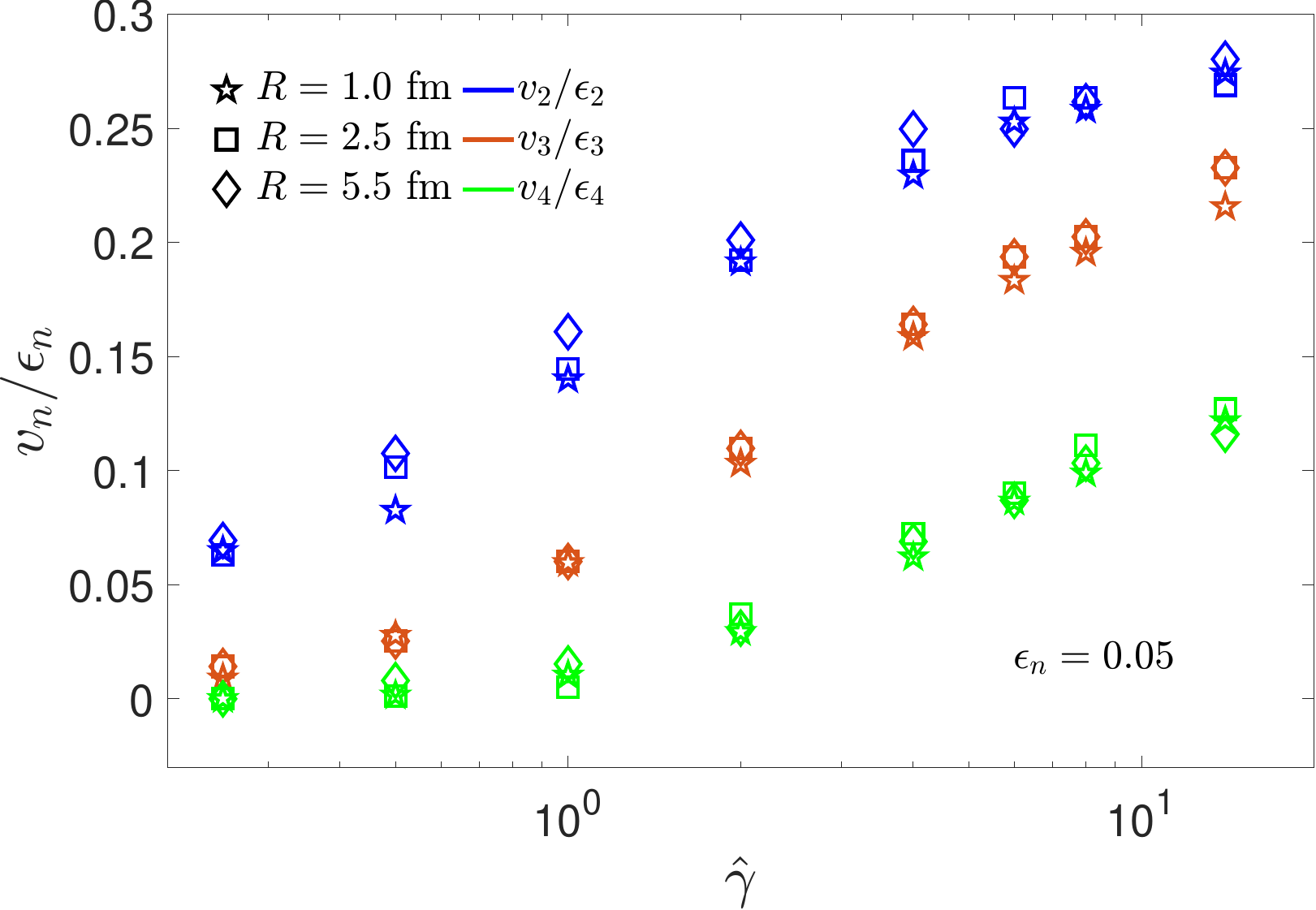}
    \caption{Integrated $v_n/\epsilon_n$ as a function of the opacity $\hat \gamma$ in the limit $t\to \infty$. Different colours correspond to different harmonics, while different point styles to different radii.}
    \label{fig:response_comparison}
\end{figure}
Finally, in Fig. \ref{fig:response_comparison} we summarise our results in the limit $t\gg R$, corresponding to the saturation value of the $v_n/\epsilon_n$ for a much wider range of opacity values.
In the Figure we show, for each value of the opacity, three different points corresponding to the different radii used above. All the values are obtained with an initial distribution $Y=\eta_s$ and $\epsilon_n=0.05$. We see that, for the range of opacity values here considered, the response is dependent almost completely on $\hat \gamma$. The spreading for different $R$ values is quite small, being $<10\%$, except for the quasi-free-streaming cases of very small opacity ($\hat\gamma <1$). It is also possible to see that the ordering in $R$ previously observed is well respected.\\
{One may wonder whether there is a dependence also on $\xi_0$. We checked that these results are independent on the initial parameter $\xi_0$, which determines the initial pressure anisotropy $P_L/P_T$, once the initial transverse energy density ${dE^0_\perp}/{d\eta}$ is fixed, as one can see from Eq. (\ref{eq:opacity}).}

\subsection{Dissipation of initial $v_2$}

\begin{figure}
    \centering
    \includegraphics[width=.515\linewidth]{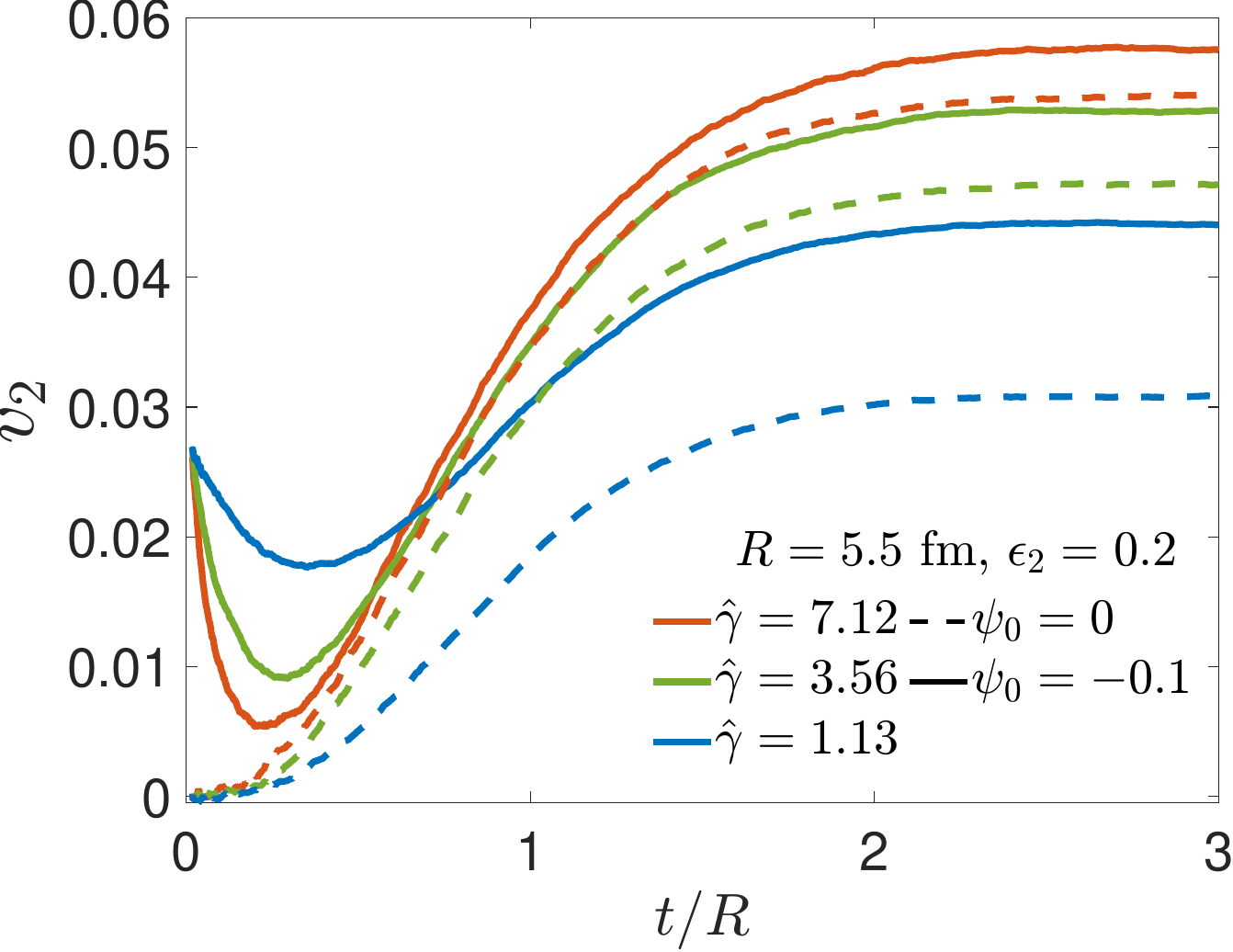}
    \includegraphics[width=.445\linewidth]{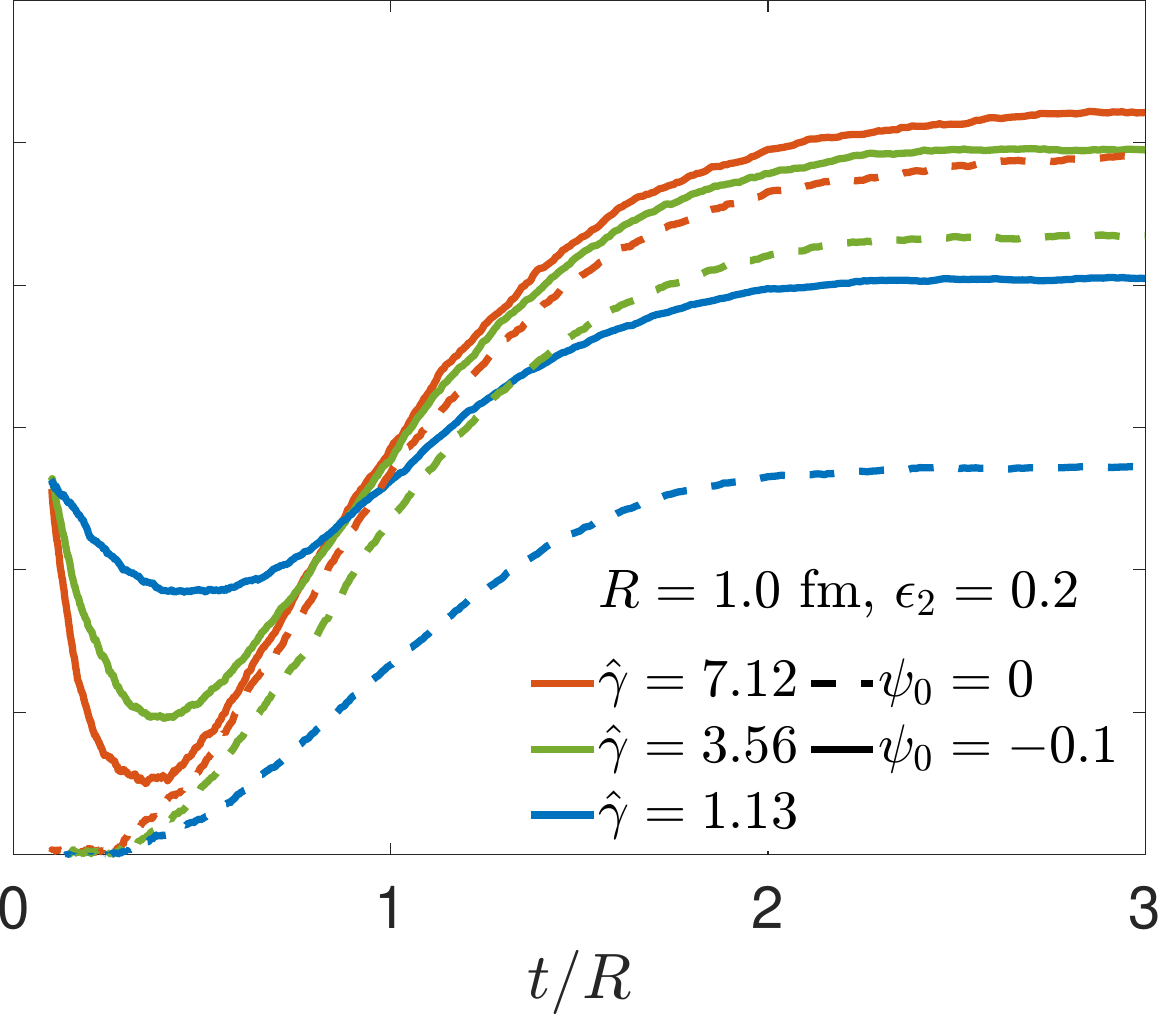}
    
    \caption{ Elliptic flow $v_2$ as a function of scaled time $t/R$ for $R=5.5$ fm (left panel) and $R=1.0$ fm (right panel), with initial eccentricity $\epsilon_2=0.2$ and two different initial conditions in the momentum space $\psi_0=-0.1$ (solid lines) and $\psi_0=0$ (dashed lines). Different colours refer to different opacity $\hat \gamma$; the corresponding $\eta/s$ values can be found in Table \ref{tb:opacity_values}.}
    \label{fig:vn_dissipation}
\end{figure}

 From now on, we will consider also non-azimuthally symmetric initial distributions in the momentum space, in order to mimic an initial correlation in momentum space typical of CGC initial conditions \cite{Krasnitz:2002ng, Schenke:2015aqa, Lappi:2015vta,Mantysaari:2017cni, Schenke:2019pmk}: in our simulation the system starts its evolution with a non zero elliptic flow $v_2(\tau_0)$. This can be easily achieved by fixing $\psi_0 \ne 0$ in Eq. \eqref{eq:modifiedRS}: we choose $\psi_0=-0.1$, which provides an initial $v_2(\tau_0)\approx 0.025$, quite similar to what found by CGC calculations \cite{Greif:2017bnr}.
 Firstly, we are interested in studying the time evolution of this initial elliptic flow for different cases, to investigate whether and how the system loses memory about these correlations in momentum space. In Fig. \ref{fig:vn_dissipation}, we consider two different system transverse sizes: $R=5.5$ fm in the left panel, which corresponds to a typical $AA$ collision, and $R=1$ in the right one, resembling small collision systems; furthermore, we explore three values of the opacity $\hat \gamma = [7.12, 3.56, 1.13]$ (the corresponding $\eta/s$ values for the two radii can be found in Table \ref{tb:opacity_values}).The initial eccentricity in coordinate space is $\epsilon_2=0.2$. Firstly, we observe that, for fixed $\hat \gamma$, the behaviour of the elliptic flow is quite independent of the transverse size: we checked this also for intermediate values of $R$. We see that the initial elliptic flow is strongly dissipated in the early times of the evolution at about {$t\approx 0.3 - 0.5$ fm}, depending on the opacity class considered. When the $v_2$ production rate, which converts the initial anisotropy in coordinate space $\epsilon_2(\tau_0)$ to the momentum space, exceeds the $v_2$ dissipation rate, the curve starts to rise, and finally at $t/R\approx 2$ the system begins to decouple and the $v_2$ saturates. We notice that in the high opacity case ($\hat \gamma=7.12)$ the system lose almost completely memory of the initial anisotropy, as one can see by comparing the final $v_2$ to the one obtained with $\psi_0=0$, that is for isotropic initial conditions in momentum space. This corresponds to $4\pi\eta/s=1-2$ for $R=5.5$ fm, as estimated for QGP, while for $R=1$ fm one needs to go down to $4\pi\eta/s\approx 0.5$ to have the same $\hat \gamma$. Even though the competition between the two processes of $v_2$ dissipation and production is present also for smaller opacity, for $\hat \gamma=3.56$ and even more for $\hat \gamma=1.18$ the system keeps memory of the initial anisotropy, with a final $v_2$ sensitively different for different $\psi_0$ cases. For instance, in a system with $R=1$ fm and $4\pi\eta/s = 1$, there is an impact on the final integrated $v_2$ which is $\gtrsim 15\%$, which goes up to $30-40\%$ for $4\pi\eta/s\approx 3$.  Not surprisingly, therefore, we find that, for the range of $\eta/s$ supposed to be explored in QGP, smaller size systems can be sensitive to initial correlations in momentum space.  It would be extremely interesting to analyse how these correlations could impact also on the $v_2(p_T)$, including realistic initial and freeze-out conditions: we plan the investigation of this aspect for future works. 
 
\subsection{Attractors in $v_n/v_{n,eq }$}

\begin{figure*}[t]
    \centering
    \includegraphics[width=.35\linewidth]{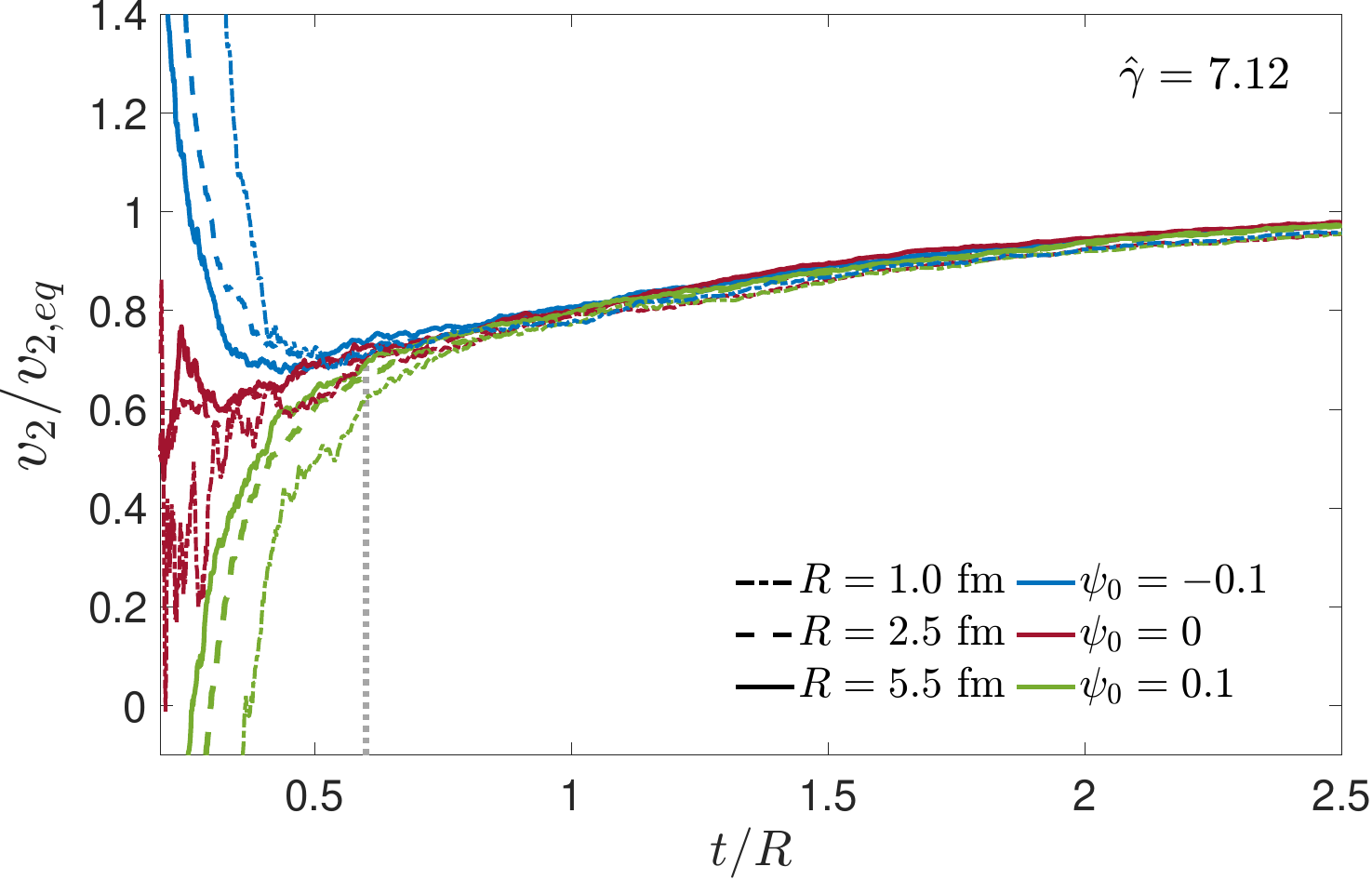}
    \includegraphics[width=.31\linewidth]{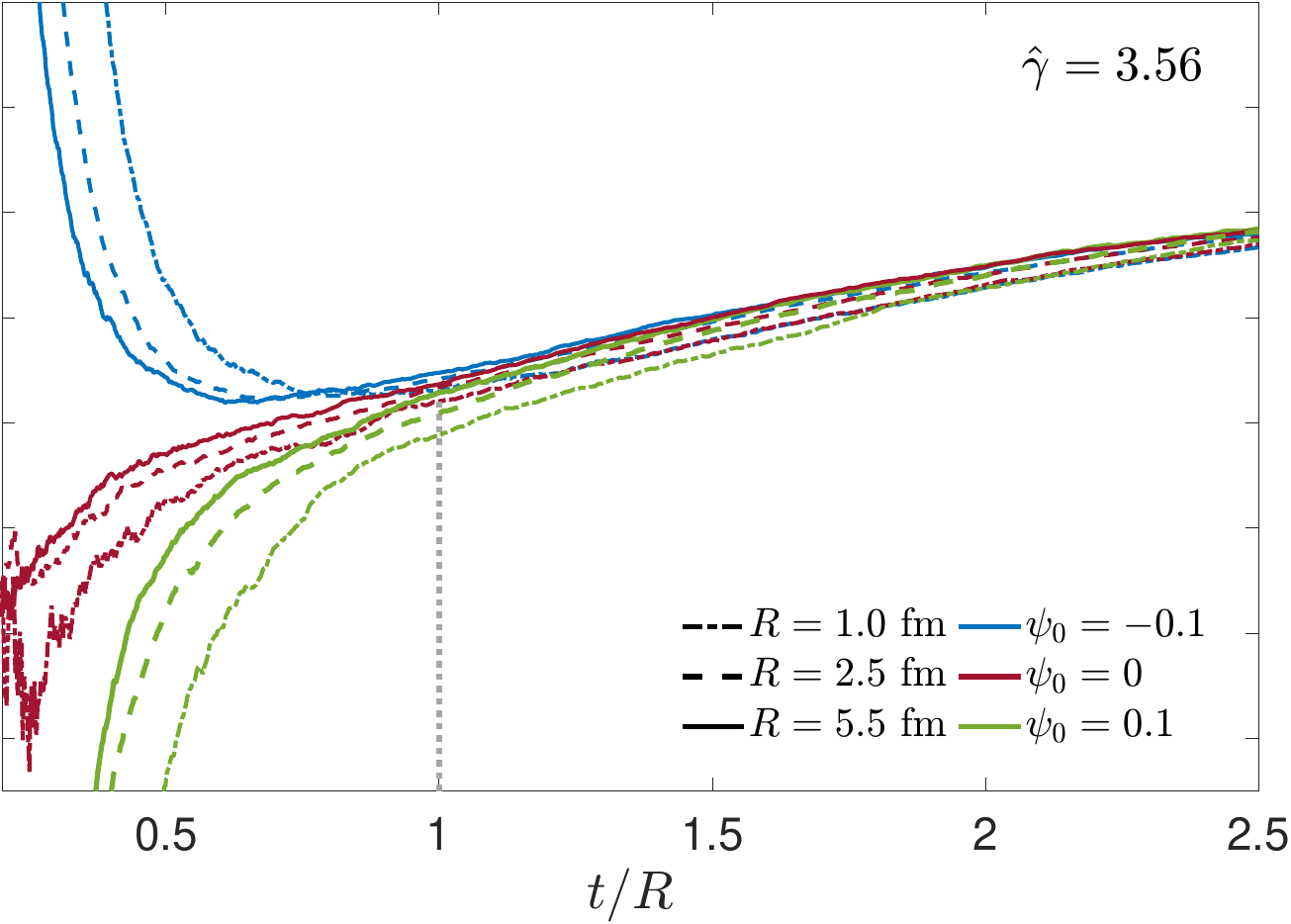}
    \includegraphics[width=.31\linewidth]{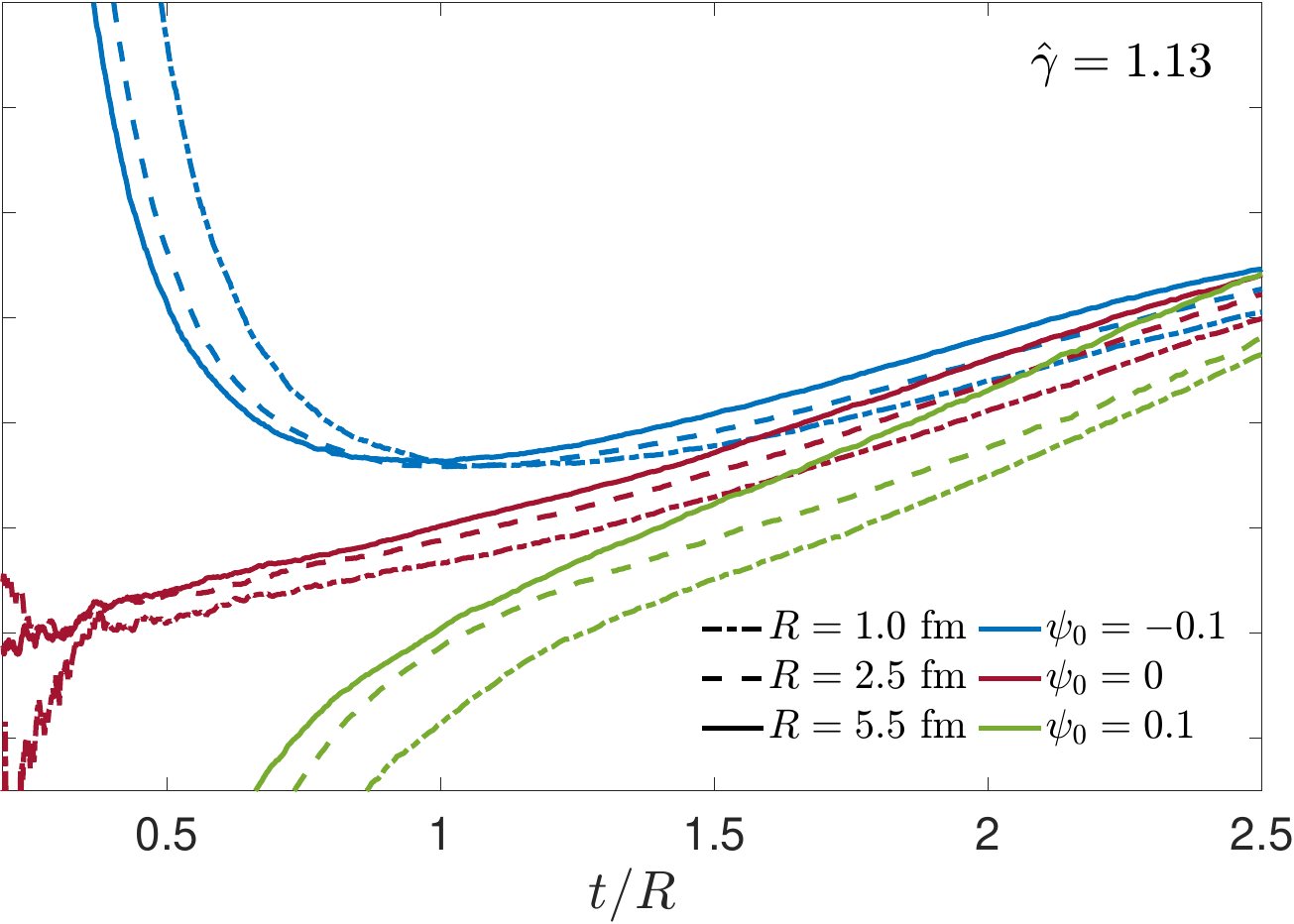}
    \caption{Normalised elliptic flow $v_2/v_{2,eq}$ for three different universality classes (from left to right) $\hat \gamma = [7.12, 3.56, 1.13]$. Different colours correspond to the three initial $\psi_0=[-0.1, 0, 0.1]$, while different line styles to the three radii $R=[1.0, 2.5, 5.5]$ fm. The corresponding $\eta/s$ values can be found in Table \ref{tb:opacity_values}.}
    \label{fig:v2veq}
\end{figure*}

We are interested in quantifying how much the computed anisotropic flows deviate from those obtained by assuming a local thermal distribution function $f(x,p)\sim \Gamma(x)\exp( - p_\mu\cdot u^\mu (x) /T(x) )$, with $u^\mu(x)$, $\Gamma(x)$ and $T(x)$ extracted locally in space and time from the simulation. The anisotropic flows computed at the equilibrium $v_n^{eq}$ at midrapidity are given by:
\begin{equation}\label{eq:equilibrium_flows}
    v_n^{eq} = \frac{\displaystyle\int d^2\mathbf x_\perp \displaystyle\int \dfrac{d^3\mathbf p}{(2\pi)^3} \cos(n\phi) \Gamma(x_\perp) \exp\left( -\frac{ p_\mu\cdot u^\mu (\mathbf x_\perp)}{T(\mathbf x_\perp)} \right)}{\displaystyle\int d^2\mathbf x_\perp \displaystyle\int \dfrac{d^3\mathbf p}{(2\pi)^3} \Gamma(x_\perp)\exp\left( -\frac{ p_\mu\cdot u^\mu (\mathbf x_\perp)}{T(\mathbf x_\perp)} \right)}.
\end{equation}

Following what has been done for the momentum moments of the distribution function, we define the normalised anisotropic flows $\overline v_n$:
\begin{equation*}
    \overline v_n (t) = \frac{v_n (t)}{v_n^{eq}(t)}.
\end{equation*}
When $t\to t_0$, since we have not introduced an initial flow $u_0^\mu=0$, $v_n^{eq}(t_0)=0$. At the same time, the initial $v_n$ is vanishing when we start with an azimuthally isotropic distribution, therefore the limit $\lim_{t\to t_0} \overline{v}_n$ is not well-defined and we have some numerical instabilities; on the other hand, if $v_n(t_0)\ne 0$ (due to $\psi_0\ne 0$), since still $v_n^{eq}(t_0)=0$, $\lim_{t\to t_0} |\overline v^n|=\infty$.
In the limit $t\to \infty$, if the system reaches the thermalisation, we expect $\lim_{t\to \infty} \overline v_n=1$ for all the cases studied.\\
We want to study whether, similarly to what shown for the momentum moments of the distribution function, there exist universality classes in terms of the opacity $\hat \gamma$ in the evolution of the normalised anisotropic flows with attractors also for the $v_2$, seeing also if the anisotropic flows developed by the medium reach those expected assuming local equilibrium. In fact $\overline v_n$ quantifies the deviation of the integrated anisotropic flow with respect to the one assuming a thermal distribution with the same $T(x), \Gamma(x)$ and $u^\mu(x)$. \\
In Fig. \ref{fig:v2veq}, we show three different plots of $\overline v_2$ in terms of $t/R$ for three different values of opacity $\hat \gamma=[7.12, 3.56, 1.13]$ (left, middle and right panel respectively). The curves correspond to different systems with three radii $R=[1.0, 2.5, 5.5]$ fm (dot-dashed, dashed and solid line respectively) and different initial momentum anisotropies $\psi_0 = [-0.1, 0, 0.1]$ (blue, dark red and green lines respectively); see Table \ref{tb:opacity_values} for the corresponding $\eta/s$. We can see that for the two larger $\hat \gamma$ the plots are qualitatively similar: all the curves, irrespectively of the transverse size, $\eta/s$ and initial $v_2(t_0)$, converge to the same behaviour when $\overline v_2 \sim 0.6 -0.7$ and then saturate to 1. We observe that the scaled time $\bar t= t/R$ at which all the curves approach the attractor depends on the opacity class considered, in particular, the smaller the opacity, the later the system will reach the universal curve: specifically, $\bar t\approx0.6$ for $\hat \gamma=7.12$ and $\bar t \approx 1$ for $\hat \gamma=3.56$. For these cases, at $t=2R$, i.e. when we expect that the system is almost completely decoupled (see Sec. \ref{sec:moments}), $v_2\approx 0.8 - 0.9\, v_{2,eq}$. Notice that the time scale at which this attractor is reached is larger than the one characterising the moments: this is due to the fact that the development of anisotropic flows is related to the transverse expansion, whose typical time scale is $R$, and not the initial longitudinal expansion.\\
As far as the small-opacity scenario ($\hat \gamma=1.13$) is concerned, we observe that the attractor behaviour seems to be partially broken, since different curves at $t=2R$ converge within a band of width $\sim 15 \%$ to a value $\bar v_2 \approx 0.8\pm 1$. This means that these systems thermalise more slowly and with different trends depending on the initial conditions, $\psi_0, R$ and $\eta/s$. According to Ref.\cite{Kurkela:2019kip}, the small opacity regime corresponds to the particle-like behaviour: indeed we see that the final collective flow is sensitively different with respect to that generated by a locally equilibrated system.

\section{Conclusions}
\label{sec:conclusions}

In this paper, we have studied how the universal behaviour that has been previously investigated in 0+1D and 1+1D systems evolves by moving to full 3+1D simulations. The results shown in this paper have been obtained by developing a transport code which numerically solves the Relativistic Boltzmann Equation with the full collision integral. We have explored different system sizes and $\eta/s$ ranging from conditions typical of $pp$ and $pA$ to Pb-Pb at top LHC energy. 
We found that, with respect to the 1D case in which the only relevant physical scale is the mean free path $\lambda_{\text{mfp}} \propto \tau_{eq}$, here a second physical quantity emerges, which is related to the transverse length of the system $R$. We observed that all the explored systems can be grouped in universality classes on the basis of the behaviour of the ratio $R/\langle\lambda_{\text{mfp}}\rangle$ for the whole studied evolution of the fireball. All the systems belonging to the same universality class share the same behaviour in different aspects, such as the approach to equilibrium of the momentum moments of the distribution function, the evolution of the Reynolds number, resembling the attractor behaviour observed in the 1D case, and the development of the collective flows $v_n$, even though the universality is partially broken for the latter. These universality classes can be directly related to the opacity parameter $\hat \gamma$ introduced in the context of RTA and ITA; indeed, our results show that at $t\sim R$ the value of $R/\langle\lambda_{\text{mfp}}\rangle$. We found that systems within the same universality class (i.e. sharing the same $\hat \gamma$) show the presence of forward and pull-back attractors in the normalised moments $\overline{\mathcal{M}}^{nm}(t,x_\perp)$ of the distribution function and also of the inverse Reynolds number $\text{Re}^{-1}(t, x_\perp)$ which directly quantifies the deviation of the system from equilibrium, therefore showing that systems with different initial conditions ($\tau_0, T_0$), size $R$ and interaction strength $\eta/s$ follow the same pattern in approaching equilibrium, provided $\hat \gamma = \tau_0 T_0/(\eta/s) (R/\tau_0)^{3/4}$ has the same value. In the limit of large $\hat \gamma$, the evolution is very close to the 1D case, while for finite $\hat \gamma$ it is possible to observe the deviation from the latter when the transverse flow starts to play a significant role ($\beta_\perp \sim 0.5$), which occurs at $t\approx R$. Note that, as observed in 1D, the presence of these attractors is due to the strong longitudinal expansion taking place in the very early stage of the evolution, at $t< R$. 
We extend this analysis also to anisotropic flows, including initial azimuthal anisotropies in coordinate space $\epsilon_n$, in order to develop corresponding anisotropic flows $v_n$ in momentum space.
We find, in agreement with what previously studied in literature in RTA and ITA, that systems with the same opacity exhibit a quite similar response curve $v_n/\epsilon_n$; however, we find that, fixing $\hat \gamma$, the response function $v_n/\epsilon_n$ is not fully universal but is a monotonic increasing function of $R$ (or equivalently $\eta/s$) resulting in a splitting up to $15\%$ going from $R=1$ fm to $R=5.5$ fm.\\
Motivated by CGC calculations that predict a finite initial anisotropy in momentum space, we have studied also how a system is able to dissipate an initial $v_2$ and what is the impact of the initial anisotropic correlation on the final $v_2$. This has been done for different values of $\hat \gamma$ and $R$, confirming once more a quite good scaling in terms of the the opacity: simulations with different transverse radius but same opacity differ by a spreading $<10\%$. More in detail, we found that for a typical scenario of $AA$ collision ($R=5.5$ fm $4\pi\eta/s\approx 1-2$) the final $v_2$ keeps little memory of the initial correlation, while the same cannot be said for smaller size systems ($R=1$ fm), in which the impact of the initial anisotropy on the final $v_2$ can be significant ($>15\%$).\\
In order to investigate the deviation of the produced $v_2$ from a thermal production, we have introduced the normalised $\overline v_n=v_n/v_{n,eq}$, following what has been done for the momentum moments. We found that systems sharing the same opacity $\hat \gamma$, show a universal behaviour in terms of $\overline v_n$ starting from $t<R$. However, already for $\hat \gamma\approx 1$ the attractor is almost completely lost, keeping a sensitive dependence on size $R$, interaction $\eta/s$ and initial anisotropy $\psi_0$. This suggests that in small systems like $pp$ for which $\hat \gamma \sim 1$ at $t\sim 2R$, which is about the decoupling time, the approach towards the thermal $v_2$ is sensitive to the details of the collision systems and to the initial conditions. In particular, the effect of initial anisotropy in momentum space is still visible even at time $t \sim 2R$, indicating that the final $v_2$ is not fully thermal, and can also keep information from the early stages of the collision. However, for typical uncertainties in the initial conditions, one can see that the impact on the final $v_2$ is up to 10\% which is quite moderate. This is no longer valid for very small $\hat \gamma \approx 1$, for which the spreading can be of the order of $\sim 20\%$.\\
For a more direct comparison with realistic collision systems, according to \cite{Kurkela:2019kip, Ambrus:2022koq} one can try to relate ranges of opacity $\hat \gamma$ values to physical systems. More in detail, $\hat \gamma\approx 5 - 10$ corresponds to Pb-Pb at 5.02 TeV and ($30-40 \, \%$) centrality class or to Au-Au collisions at 200 GeV and ($20-30 \, \%$) centrality class while $\hat \gamma\approx 2 - 4$ corresponds to $p$-Pb at 5.02 TeV central collision and finally  $\hat \gamma\approx 1 - 3$ corresponds to $pp$ collisions at 5.02 TeV. From the study of the response functions $v_n/\epsilon_n$  we observe that higher order harmonic flows permit a larger spreading of the  $v_n/\epsilon_n$ for different $\hat \gamma$ in the range of values explored; while from the investigation of $v_2/v_{2, eq}$ we can conclude that smaller $\hat \gamma$ systems, such as $p$-Pb or $pp$, do not have enough time to wash out the information about the initial conditions, especially in momentum space.\\
Finally, according to our results, the opacity parameter $\hat \gamma$ can be considered a quite good label for universality classes for $\hat \gamma >1$, since the deviation within a sensitive range of transverse dimension $R$ and specific viscosity $\eta/s$ has been found to be $\lesssim 10\%$, both in the response functions and in the $v_2/v_{2,eq}$, even when initial correlations in momentum space have been introduced. Such spreading, however, gets even smaller for higher values of $\hat \gamma$, which are typical of systems with large size ($pA$ or $AA$). For $\gamma\lesssim 1$, instead, the universality is broken more significantly, with variations that reach $20\%$; this case could be of interest for $pp$ collisions with $4\pi\eta/s \sim 2-3$.

\subsection*{Acknowledgments}
We acknowledge the funding from UniCT under PIACERI ‘Linea di intervento 1’ (M@uRHIC).
This work is (partially) supported by ICSC – Centro Nazionale di Ricerca in High Performance Computing, Big Data and Quantum Computing, funded by European Union – NextGenerationEU.

\subsection*{Data Availability Statement}
The raw data of the figures are available in the Zenodo repository: \url{https://doi.org/10.5281/zenodo.13829380}.

\end{document}